\documentstyle[12pt,axodraw]{article}

\def\pd{\partial}
\def\a{\alpha}
\def\b{\beta}
\def\dl{\delta}
\def\s{\sigma}

\def\vphi{\varphi}
\def\eps{\epsilon}
\def\veps{\varepsilon}
\def\lam{\lambda}

\def\hg{{\hat g}}
\def\hnabla{{\hat \nabla}}
\def\hR{{\hat R}}
\def\hE{{\hat E}}
\def\hDelta{{\hat \Delta}}

\def\bx{\bar{x}}
\def\by{\bar{y}}
\def\bz{\bar{z}}
\def\bw{\bar{w}}
\def\bu{\bar{u}}
\def\bv{\bar{v}}
\def\prm{m^{\prime}}

\def\prs{s^{\prime}}

\def\tvphi{\tilde{\varphi}}
\def\tq{\tilde{q}}
\def\ta{\tilde{a}}
\def\tb{\tilde{b}}
\def\tc{\tilde{c}}
\def\td{\tilde{d}}
\def\te{\tilde{e}}
\def\bfc{{\bf c}}
\def\bfb{{\bf b}}
\def\S{{\rm S}}
\def\A{{\rm A}}

\def\gm{\gamma}
\def\hgm{{\hat \gamma}}

\def\Om{\Omega}
\def\sq{\sqrt}
\def\e{\hbox{\large \it e}}
\def\half{\frac{1}{2}}
\def\fr{\frac}
\def\pp{\prime}

\def\C{{\bf C}}
\def\D{{\bf D}}

\def\E{{\bf E}}
\def\F{{\bf F}}
\def\G{{\bf G}}
\def\H{{\bf H}}
\def\I{{\bf I}}
\def\V{{\bf V}}
\def\U{{\bf U}}

\def\oE{\overline{\E}}
\def\oH{\overline{\H}}
\def\oD{\overline{\D}}
\def\oG{\overline{\G}}
\def\oI{\overline{\I}}
\def\bb{\begin{equation}}
\def\ee{\end{equation}}
\def\bba{\begin{eqnarray}}
\def\eea{\end{eqnarray}}

\begin{document}

\begin{titlepage}

\begin{tabbing}
   qqqqqqqqqqqqqqqqqqqqqqqqqqqqqqqqqqqqqqqqqqqqqq
   \= qqqqqqqqqqqqq  \kill
         \>  {\sc KEK-TH-940}    \\
%         \>       hep-th/ \\
         \>  {\sc February 2004}

\end{tabbing}

\vspace{5mm}

\begin{center}
{\Large {\bf Building Blocks of Physical States in a Non-Critical 3-Brane 
on $R \times S^3$}}
\end{center}

\vspace{5mm}

\begin{center}
{\sc Ken-ji Hamada}
\end{center}

\begin{center}
{\it Institute of Particle and Nuclear Studies, \break
High Energy Accelerator Research Organization (KEK),} \\
{\it Tsukuba, Ibaraki 305-0801, Japan}
\end{center}

\vspace{7mm}

\begin{abstract}
The physical states in a world-volume model of a non-critical 3-brane 
are systematically constructed using techniques of four-dimensional 
conformal field theories on $R \times S^3$ developed recently. 
Invariant combinations of creation modes under a special conformal 
transformation provide building blocks of physical states. 
Any state can be created by acting with such building blocks 
on a conformally invariant vacuum 
in an invariant way under the other conformal charges: the Hamiltonian 
and rotation generators on $S^3$.
We explicitly construct building blocks for scalar, vector and gravitational 
fields, and classify them as finite types.
\end{abstract}
\end{titlepage}

\section{Introduction}
\setcounter{equation}{0}
\noindent

Conformal invariance is one of the most important symmetry in statistical mechanics,  
strings and quantized gravity. Applications of conformally invariant quantum 
field theories to such physics are modern streams to study their non-perturbative effects. 
Especially, it is well-known that in two dimensions an infinite number of generators form 
the Virasoro algebra, and yield powerful constraints on the classification and 
the physical properties of two-dimensional critical points~\cite{isz} and strings~\cite{gsw,kpz}.
In higher dimensions, the number of the generators becomes finite, but the conformal 
invariance is still powerful.

Recently, there have been remarkable developments in four-dimensional conformal 
field theories~\cite{cardy,riegert,amm97,hamada02,hh,maldacena,nst,fp}.   
As an advance on spacetime physics~\cite{riegert,amm97,hamada02,hh}, 
a renormalizable world-volume model of a non-critical 3-brane~\cite{hamada02,hh} was quantized 
on $R\times S^3$~\cite{amm97,hh} in a strong gravity phase, in which world-volume 
fluctuations are dominated by the conformal field. 
This model possesses exact conformal symmetry as a realization of the background-metric 
independence. The conformal symmetry is generated by 15 conformal charges:    
the Hamiltonian, $H$, the rotation generators on $S^3$, $R_{MN}$, and the charges 
for the special conformal transformations, $Q_M$, and their conjugates, $Q_M^\dag$. 
These charges satisfy the conformal algebra~\cite{amm97,hh}:
\bba
     \left[ Q_M, Q^{\dag}_N \right] &=& 2\dl_{MN} H + 2R_{MN},
           \nonumber \\
    \left[ H, Q_M \right] &=& -Q_M,
           \nonumber \\
    \left[ H, R_{MN} \right] &=& \left[ Q_M, Q_N \right] = 0,
           \nonumber  \\
    \left[ Q_M, R_{M_1 M_2} \right] &=& \dl_{M M_2}Q_{M_1}
                 -\eps_{M_1}\eps_{M_2}\dl_{M -M_1}Q_{-M_2} ,
            \nonumber  \\
    \left[ R_{M_1 M_2}, R_{M_3 M_4} \right]
        &=& \dl_{M_1 M_4} R_{M_3 M_2} -\eps_{M_1}\eps_{M_2} \dl_{-M_2 M_4} R_{M_3 -M_1}
            \nonumber \\
    && - \dl_{M_2 M_3} R_{M_1 M_4} +\eps_{M_1}\eps_{M_2} \dl_{-M_1 M_3} R_{-M_2 M_4} .
            \label{algebra}
\eea
Explicit forms of the conformal charges for scalar~\cite{amm97}, 
vector~\cite{hh} and gravitational fields~\cite{hh,amm97} 
have been constructed.

The conformal invariance imposes strong constraints on the physical states. 
The physical states in a non-critical 3-brane must satisfy the conformal invariance  
conditions,\footnote{%%%%%%%%(footnote)%%%%%%
Here, these charges include the ghost sector in the radiation$^+$ gauge discussed 
in Appendix F. 
}%%%%%%%%%%%%%% 
\bb
    Q_M | {\rm phys}\rangle = 
     H | {\rm phys}\rangle = R_{MN} | {\rm phys}\rangle =0. 
          \label{phys-conditions}
\ee
Such states have been constructed in Ref.\cite{hh}. In this paper, we further develop the arguments 
and give a systematic method to construct and classify the physical states.

The physical state is divided into sectors of matter fields  and  gravitational fields. 
Here, we consider the scalar and vector fields as matter fields. The gravitational fields are 
further decomposed into two sectors: the conformal mode and the traceless mode. 
Dynamical fields are mode-expanded in the spherical tensor harmonics on $S^3$. 
The standard Fock state created by acting with a creation mode on a vacuum    
is an eigenstate of the Hamiltonian belonging to a certain representation 
of the rotation group on $S^3$.     
However, the charge for the special conformal transformation, $Q_M$, maps 
a creation mode to another creation mode belonging to a different representation.   
Therefore, this charge yields a strong constraint on the physical states.
We seek a $Q_M$-invariant combination of creation modes in each sector.  
Such an operator provides a building block of physical states. 
We have found that apart from a few creation modes, such a building block is obtained 
by a particular combination of the products of two creation modes, 
and classified as finite types.
Any state satisfying the first conditon in (\ref{phys-conditions})
is created by acting with building blocks on a conformally invariant vacuum.   
The third condition in (\ref{phys-conditions}) can be easily satisfied by 
combining all sectors in a rotation-invariant way. 
The Hamiltonian condition is imposed last by adjusting the zero-mode momentum 
existing in the conformal field.
In this way we can obtain an infinite number of physical states in a non-critical 
3-brane.

This paper is organized as follows. In the next section we briefly review a world-volume 
model of a non-critical 3-brane on $R \times S^3$ developed in Ref.\cite{hh}. 
We here give mode expansions of 
dynamical fields, canonical commutation relations, and explicit forms of the conformal 
charges and so on. In Sect.3 we discuss conformally coupled scalar fields  
in detail.  Basic ideas on how to construct and classify physical states in a non-critical 
3-brane are given here. 
A $Q_M$-invariant creation operator with a scalar index is the building block 
in this sector. 
In order to help our intuitive understanding of the structure of states, 
a graphical representation is introduced here. 
In Sect.4 we develop an argument for the case of vector fields. 
We find that there are three types of building blocks with scalar, vector and 
rank 2 tensor indices.
In Sect.5 we construct building blocks for the traceless mode sector. 
For gravitational fields, both positive-metric and negative-metric modes 
are required to form a closed conformal algebra. However, 
no negative-metric modes commute with the charge $Q_M$, and thus these modes 
are not independent physical modes.   
We find that $Q_M$-invariant building blocks are given by particular combinations 
of the products of such positive-metric and negative-metric modes, 
apart from the lowest positive-metric mode.   
They are classified as seven types with tensor indices up to rank 4.  
The building blocks for the conformal mode sector are constructed in Sect.6. 
They are classified as two types with a scalar index. 
Because of the existence of the zero mode, the conformal mode sector is managed separately. 
Physical states are constructed in Sect.7 by acting with building blocks on a conformally 
invariant vacuum in such a way that the conditions for the Hamiltonian and the rotation generators 
in Eqs.(\ref{phys-conditions}) are satisfied. 
The physical states up to level 6 are constructed explicitly. 
Sect.8 is devoted to conclusions and a discussion.

\section{A World-Volume Model of a Non-Critical 3-Brane}
\setcounter{equation}{0}
\noindent

In this section we review recent developments on a world-volume model 
of a non-critical 3-brane~\cite{hh}.

\subsection{Canonical Quantization on $R\times S^3$}

The world-volume metric is decomposed to the conformal mode, $\phi$, and 
the traceless mode, $h^{\lam}_{~\nu}$, as
\bb
      g_{\mu\nu} =\e^{2\phi} \hg_{\mu\lam}
                   (\dl^{\lam}_{~\nu} +t h^{\lam}_{~\nu} + \cdots ),
\ee
with $tr(h)=0$, and $\hg_{\mu\nu}$ is the background metric. In this paper we use 
the $R\times S^3$ background with the Lorentzian signature $(-1,1,1,1)$. 
The traceless mode, whose field strength is given by the squre of the Weyl tensor 
divided by $t^2$, is handled perturbatively in terms of the coupling $t$, while 
the conformal mode is treated {\it exactly}.  
Recently, it was shown that 
the pertubative expansion in $t$ is renormalizable and asymptotically free~\cite{hamada02}.
The asymptotic freedom implied at very high energies above the Planck mass 
the Weyl tensor should vanish, and world-volume fluctuations 
are dominated by the conformal mode. The dynamics is described by a four-dimensional conformal field theory (CFT$_4$). 
In the following we consider the case of the vanishing coupling 
realized at very high energies where mass scales can be neglected.

   The action for the conformal mode is induced from the measure, 
as in the case of a non-critical string~\cite{kpz}. The four-dimensional counteraction 
of the Polyakov-Liouville action in a non-critical string is 
the Riegert action~\cite{riegert}, 
given by
\bb
   S=-\fr{b_1}{(4\pi)^2} \int d^4 x \sq{-\hg} \left\{
       2 \phi \hDelta_4 \phi + \hE_4 \phi \right\},
\ee
where $\sq{-g}\Delta_4$ is a conformally invariant 4-th order operator,
defined by $\Delta_4 = \Box^2 +2 R^{\mu\nu}\nabla_{\mu}\nabla_{\nu}-\fr{2}{3}R\Box
+ \fr{1}{3} (\nabla^{\mu}R)\nabla_{\mu}$, 
and $E_4 = G_4 -\fr{2}{3}\Box R$ and $G_4$ is the Euler density. 
This action is related to the conformal anomaly proportional to the Euler density.  
The coefficient $b_1$ has been calculated as
$b_1=\fr{1}{360} \left(N_X +\fr{11 N_W}{2} +62 N_A \right) + \fr{769}{180}$,
where $N_X$, $N_W$ and $N_A$ are the number of conformal scalar fields, 
Weyl fermions and gauge fields, respectively.

    The kinetic term of the traceless mode is given by the linearized form of
the Weyl action, which is invariant under a gauge transformation,
$\dl_\xi h_{\mu\nu}=\hnabla_\mu \xi_\nu +\hnabla_\nu \xi_\mu
-\fr{1}{2}\hg_{\mu\nu}\hnabla^\lam \xi_\lam$.\footnote{%%%%%%%(footnote)%%%%%%%
More precisely, the diffeomorphism transformation, 
$\dl_\xi g_{\mu\nu}=g_{\nu\lam}\nabla_\mu \xi^\lam +g_{\mu\lam}\nabla_\nu \xi^\lam$, 
is decomposed into the transformation of the conformal mode, 
$\dl_\xi \phi =\xi^\lam \pd_\lam \phi +\fr{1}{4} \hnabla_\lam \xi^\lam$,  
and that of the traceless mode expanded by the coupling as
\bba
  \dl_{\xi}h_{\mu\nu} 
  &=& \fr{1}{t} \left(\hnabla_{\mu}\xi_{\nu}+\hnabla_{\nu}\xi_{\mu}
            -\fr{1}{2}\hg_{\mu\nu}\hnabla^\lam \xi_\lam \right)
           \nonumber \\
  && + \xi^\lam \hnabla_\lam h_{\mu\nu} 
     +\half h_{\mu\lam} \left( \hnabla_\nu \xi^\lam -\hnabla^\lam \xi_\nu \right) 
     +\half h_{\nu\lam} \left( \hnabla_\mu \xi^\lam -\hnabla^\lam \xi_\mu \right) 
   +o(t),   \nonumber
\eea
where $\xi_\mu=\hg_{\mu\nu}\xi^\nu$. 
The linearized Weyl action is invariant under the lowest term of this 
transformation, $\dl_\xi h_{\mu\nu}$, and therefore we write here only this term, 
in which $\xi/t$ is replaced by $\xi$. 
It is worth commenting that in the case that $\xi^\mu$ are the conformal Killing vectors,  
the lowest term of the transformation vanishes, and thus 
the next order terms  become effective, even in the linearized model. 
This transformation is the conformal transformation discussed in the text.     
}%%%%%%%%%%%%%%%%%%
To quantize the traceless mode, we take the ${\rm radiation}^+$ gauge~\cite{hh},
\bba
   && h^0_{~0}=\hnabla_i h^i_{~0}=\hnabla_i h^i_{~j}=0 ,
       \nonumber \\
   &&  h_{\half}^{~0i}=0,
     \label{radiation}
\eea
where $h_{1/2}^{~0i}$ is the lowest mode of $h^{0i}$ in the spherical-harmonics 
expansion defined below. 
In this gauge, the space of the residual gauge symmetry becomes  equivalent to that  
spanned by conformal Killing vectors on $R\times S^3$.

   A world-volume model of a non-critical 3-brane coupled to the conformal scalar 
and vector fields is considered here. 
For the vector field, the radiation gauge, $A^0=\hnabla_i A^i=0$, is taken. 
The action on $R \times S^3$ in the radiation$^+$ gauge is given by
\bba
    I_{CFT} &=& \int dt \int_{S^3} d^3 x \sq{\hgm} \biggl\{
     -\fr{2b_1}{(4\pi)^2}\phi \left(
        \pd_t^4 -2\hnabla^2\pd_t^2 +\hnabla^4 + 4\pd_t^2 \right) \phi
                 \nonumber  \\
     && \qquad\qquad
        -\fr{1}{2}h^i_{~j} \left( \pd_t^4 -2\hnabla^2\pd_t^2 + \hnabla^4
                     + 8\pd_t^2-4\hnabla^2 +4 \right) h^j_{~i}
                \nonumber     \\
     && \qquad\qquad
       + h^0_{~i} \left( \hnabla^2+2 \right)
                 \left(-\pd_t^2 +\hnabla^2 -2 \right) h^{0i}
                \nonumber    \\
     && \qquad\qquad
       +\fr{1}{2}X \left( -\pd_t^2+\hnabla^2-1 \right) X
                  \nonumber  \\
     && \qquad\qquad
       +\fr{1}{2}A_i \left( -\pd_t^2 +\hnabla^2 -2 \right) A^i \biggr\},
            \label{CFT}
\eea
where $\hnabla^2=\hnabla^i \hnabla_i$ is the Laplacian on $S^3$. 
The background metric is parametrized as in (\ref{metric}), in which the 
radius of $S^3$ is taken to be unity.
Here and henceforth, $t$ denotes the time, not the traceless mode coupling.

   Since the model is conformally invariant, we could take any conformal background. 
The advantages that we use the $R\times S^3$ background are that mode expansions 
of higher-derivative gravitational fields have quite simple forms, 
and then canonical commutation relations of these modes become diagonal. 
Furthermore, we can use the properties of Clebsch-Gordan coefficients 
because the isometry group of $S^3$ is $SO(4)=SU(2)\times SU(2)$.

  Dynamical fields are expanded in symmetric-traceless-transverse (ST$^2$) 
spherical tensor harmonics~\cite{ro,nst,hh}.
The ST$^2$ tensor harmonics of rank $n$ are classified using
the $(J+\veps_n,J-\veps_n)$ representation of $SU(2) \times SU(2)$ 
for each sign of $\veps_n=\pm \fr{n}{2}$. 
They are, denoted by $Y^{i_1 \cdots i_n}_{J (M \veps_n)}$,  
the eigenfunction of the Laplacian on $S^3$,
\bb
     \hnabla^2 Y^{i_1 \cdots i_n}_{J (M \veps_n)}
      =\{ -2J(2J+2)+n \} Y^{i_1 \cdots i_n}_{J (M \veps_n)},
\ee
where $J ~(\geq \fr{n}{2})$ takes integer or half-integer values,
and  $M=(m,\prm)$ represents the multiplicity for $\veps_n$, which takes the 
following values:
\bba
  m &=&-J-\veps_n,~ -J-\veps_n+1, \cdots, J+\veps_n-1,~ J+\veps_n,
             \nonumber \\
  \prm &=&-J+\veps_n,~ -J+\veps_n+1, \cdots, J-\veps_n-1,~ J-\veps_n.
\eea
Thus, the multiplicity of the ST$^2$ tensor harmonic of rank $n$ is
given by the product of the left and right $SU(2)$ 
multiplicities, $(2(J+\veps_n)+1)(2(J-\veps_n)+1)$, for each sign
of $\veps_n$, and thus it is totally $2(2J+n+1)(2J-n+1)$ for $n \geq 1$.
For $n=0$, the multiplicity is given by $(2J+1)^2$.
The explicit forms of the ST$^2$ tensor harmonics of any rank are given in Ref.\cite{hh}.
They are normalized as
\bb
    \int_{S^3} d\Om_3 Y^{i_1 \cdots i_n *}_{J_1 (M_1 \veps^1_n)}
                      Y_{i_1 \cdots i_n J_2 (M_2 \veps^2_n)}
     =\dl_{J_1J_2}\dl_{M_1M_2}\dl_{\eps^1_n \veps^2_n},
\ee
where 
\bb
        Y^{i_1 \cdots i_n *}_{J (M \veps_n)}
        =(-1)^n \eps_M Y^{i_1 \cdots i_n}_{J (-M \veps_n)}.
\ee 
Below, we use the following parametrizations for the tensor indices 
up to rank 4:
\bb
    \veps_0=0, \quad \veps_1=y=\pm\half, \quad \veps_2=x=\pm 1, \quad
    \veps_3=z=\pm\fr{3}{2}, \quad \veps_4=w=\pm 2. 
\ee

{}From the CFT$_4$ action (\ref{CFT}), we can easily obtain the equations of motion 
and mode expansions of the dynamical fields.
The scalar and vector fields are expanded as 
\bba
   X &=& \sum_{J \geq 0}\sum_M \fr{1}{\sq{2(2J+1)}} \left\{
            \vphi_{JM}\e^{-i(2J+1)t}Y_{JM}
            + \vphi^{\dag}_{JM}\e^{i(2J+1)t}Y^*_{JM} \right\},
                        \\
   A^i &=& \sum_{J \geq \half}\sum_{M,y} \fr{1}{\sq{2(2J+1)}} \left\{
            q_{J(My)}\e^{-i(2J+1)t}Y^i_{J(My)}
            + q^{\dag}_{J(My)}\e^{i(2J+1)t}Y^{i*}_{J(My)} \right\}.
              \nonumber  \\
       &&
\eea
These fields are normalized such that the canonical commutation relations become 
\bba
   && [\vphi_{J_1 M_1}, \vphi^{\dag}_{J_2 M_2} ]
      = \dl_{J_1 J_2}\dl_{M_1 M_2},
             \nonumber          \\
   && [q_{J_1 (M_1 y_1)}, q^{\dag}_{J_2 (M_2 y_2)} ]
      = \dl_{J_1 J_2}\dl_{M_1 M_2}\dl_{y_1 y_2},
\eea
where $\dl_{M_1 M_2}=\dl_{m_1 m_2}\dl_{\prm_1\prm_2}$.

The mode expansions of gravitational fields are given by
\bba
  \phi &=& \fr{\pi}{2\sq{b_1}} \biggl[ 2(\hat{q} +\hat{p}t) Y_{00}
                 \nonumber  \\
   && + \sum_{J \geq \half}\sum_M \fr{1}{\sq{J(2J+1)}} \left\{
            a_{JM}\e^{-i2Jt}Y_{JM}
            + a^{\dag}_{JM}\e^{i2Jt}Y^*_{JM} \right\}
                  \nonumber  \\
   && + \sum_{J \geq 0}\sum_M \fr{1}{\sq{(J+1)(2J+1)}} \left\{
            b_{JM}\e^{-i(2J+2)t}Y_{JM}
            + b^{\dag}_{JM}\e^{i(2J+2)t}Y^*_{JM} \right\} \biggr],
                   \nonumber  \\
   &&  \label{conformal-mode}  \\
  h^{ij} &=&
       \fr{1}{4} \sum_{J \geq 1}\sum_{M,x} \fr{1}{\sq{J(2J+1)}} \left\{
            c_{J(Mx)}\e^{-i2Jt}Y^{ij}_{J(Mx)}
            + c^{\dag}_{J(Mx)}\e^{i2Jt}Y^{ij*}_{J(Mx)} \right\}
              \nonumber       \\
   && + \fr{1}{4} \sum_{J \geq 1}\sum_{M,x} \fr{1}{\sq{(J+1)(2J+1)}} \Bigl\{
            d_{J(Mx)}\e^{-i(2J+2)t}Y^{ij}_{J(Mx)}
                \nonumber         \\
   && \qquad\qquad\qquad\qquad\qquad\qquad\qquad
            + d^{\dag}_{J(Mx)}\e^{i(2J+2)t}Y^{ij*}_{J(Mx)} \Bigr\},
                          \\
  h^{0i} &=&
       \fr{1}{2}\sum_{J \geq 1}\sum_{M,y} \fr{1}{\sq{(2J-1)(2J+1)(2J+3)}}
           \Bigl\{  e_{J(My)}\e^{-i(2J+1)t}Y^i_{J(My)}
                     \nonumber \\
   && \qquad\qquad\qquad\qquad\qquad\qquad\qquad
          + e^{\dag}_{J(My)}\e^{i(2J+1)t}Y^{i*}_{J(My)} \Bigr\},
\eea
where $Y_{00}=\fr{1}{\sq{{\rm Vol}(S^3)}}=\fr{1}{\sq{2}\pi}$. 
The radiation$^+$ gauge (\ref{radiation}) implies that in the mode expansion 
of $h^{0i}$ the lowest mode with $J=\half$ is removed,  because this mode satisfies  
the equation $(\hnabla^2 +2) h^{0i}_{1/2}=0$, and therefore there is no kinetic term of 
this mode.
The canonical commutation relations of the gravitational modes are given by
\bba
    && \left[ \hat{q}, \hat{p} \right] = i,
              \nonumber  \\
    && \left[a_{J_1 M_1}, a^{\dag}_{J_2 M_2} \right]
      = -\left[b_{J_1 M_1}, b^{\dag}_{J_2 M_2} \right]
      = \dl_{J_1 J_2}\dl_{M_1 M_2},
             \nonumber          \\
    && \left[c_{J_1 (M_1 x_1)}, c^{\dag}_{J_2 (M_2 x_2)} \right]
      = -\left[d_{J_1 (M_1 x_1)}, d^{\dag}_{J_2 (M_2 x_2)} \right]
      = \dl_{J_1 J_2} \dl_{M_1 M_2}\dl_{x_1 x_2},
             \nonumber   \\
    && \left[e_{J_1 (M_1 y_1)}, e^{\dag}_{J_2 (M_2 y_2)} \right]
      = -\dl_{J_1 J_2}\dl_{M_1 M_2}\dl_{y_1 y_2}.
\eea
Thus, the $a_{JM}$ and $c_{J (M x)}$ are positive-metric modes,  
and the $b_{JM}$, $d_{J (M x)}$ and $e_{J (M y)}$ are negative-metric modes.

\subsection{Conformal charges on $R \times S^3$}

    Because the conformal field, $\phi$, is quantized exactly without introducing
the coupling constant concerning this field, the model possesses exact conformal invariance.
This conformal symmetry is generated by 15 charges: the Hamiltonian, the 6 rotation 
generators on $S^3$, and the 8 charges for the special conformal transformations.

    The Hamiltonian is given by
\bba
   H&=& \half \hat{p}^2 +b_1
        +\sum_{J \geq 0}\sum_M \{ 2J a^{\dag}_{JM}a_{JM}-(2J+2)b^{\dag}_{JM}b_{JM} \}
                \nonumber  \\
    && +\sum_{J \geq 1}\sum_{M,x} \{ 2J c^{\dag}_{J(Mx)}c_{J(Mx)}
                 -(2J+2)d^{\dag}_{J(Mx)}d_{J(Mx)} \}
                \nonumber  \\
    && -\sum_{J \geq 1}\sum_{M,y} (2J+1) e^{\dag}_{J(My)}e_{J(My)}.
                \nonumber  \\
    && + \sum_{J \geq 0}\sum_M (2J+1)\vphi^{\dag}_{JM}\vphi_{JM}
                \nonumber  \\
    && +\sum_{J \geq \half}\sum_{M,y} (2J+1) q^{\dag}_{J(My)}q_{J(My)}.
\eea

The rotation generators on $S^3$, $R_{MN}$, satisfy the 
relations
\bb
     R_{MN}=-\eps_M \eps_N R_{-N-M}, \qquad  R^{\dag}_{MN}=R_{NM},  
\ee
where $\eps_M=(-1)^{m-\prm}$, and the indices, $M$ and $N$, are the ${\bf 4}$ 
vectors on $SU(2)\times SU(2)$. From these relations only $6$ of these generators 
are independent. If we parametrize the ${\bf 4}$
representation of $SU(2)\times SU(2)$ as
$\{ (\half,\half), (\half,-\half), (-\half,\half), (-\half,-\half) \}
=(1,2,3,4)$, 
and identify
$A_+ =R_{31}$, $A_- =R^{\dag}_{31}$, $A_3=\half (R_{11}+R_{22})$,
$B_+ =R_{21}$, $B_- =R^{\dag}_{21}$ and $B_3=\half (R_{11}-R_{22})$,
the closed algebra of $R_{MN}$ in (\ref{algebra}) can be expressed by 
the standard $SU(2)\times SU(2)$ algebra, i.e.,
\bba
  &&  [A_+, A_-]=2A_3, \qquad [A_3, A_{\pm}]=\pm A_{\pm},
         \nonumber \\
  &&  [B_+, B_-]=2B_3, \qquad [B_3, B_{\pm}]=\pm B_{\pm},
\eea
and $A_{\pm,3}$ and $B_{\pm,3}$ commute. 
The generators $A_{\pm,3}$ $(B_{\pm,3})$ act on the left (right) $SU(2)$ index of $M=(m,\prm)$ 
in each mode.
Explicit forms of $R_{MN}$ are given in Ref.\cite{hh}.

  The most important conformal charges to determine the physical states are the charges for the
special conformal transformations, denoted by $Q_M$, and their hermite conjugates.
For the scalar and vector fields, they are given by
\bba
    Q_M &=& \sum_{J \geq 0}\sum_{M_1, M_2} \C^{\half M}_{JM_1, J+\half M_2}
           \sq{(2J+1)(2J+2)}  \tilde{\vphi}^{\dag}_{JM_1}\vphi_{J+\half M_2} 
              \nonumber \\
    &&  -\sum_{J \geq \half}\sum_{M_1,y_1, M_2, y_2}
           \D^{\half M}_{J(M_1 y_1), J+\half (M_2 y_2)}
               \nonumber  \\
    && \qquad\qquad \times
            \sq{(2J+1)(2J+2)} \tilde{q}^{\dag}_{J(M_1 y_1)} q_{J+\half (M_2 y_2)}.
\eea
For the gravitational fields, they are given by
\bba
    Q_M &=& \left( \sq{2b_1}-i\hat{p} \right) a_{\half M}
                 \nonumber \\
        && +\sum_{J \geq 0}\sum_{M_1,M_2} \C^{\half M}_{JM_1, J+\half M_2}
              \Bigl\{ \a(J)  \tilde{a}^{\dag}_{JM_1} a_{J+\half M_2}
                  \nonumber  \\
        &&\qquad\qquad\qquad\qquad\qquad
             +\b(J) \tilde{b}^{\dag}_{JM_1} b_{J+\half M_2}
                  \nonumber  \\
        &&\qquad\qquad\qquad\qquad\qquad
             +\gm(J) \tilde{a}^{\dag}_{J+\half M_2} b_{J M_1} \Bigr\}
                  \nonumber  \\
        &&+ \sum_{J \geq 1}\sum_{M_1,x_1, M_2,x_2}
            \E^{\half M}_{J(M_1 x_1), J+\half (M_2 x_2)}
          \Bigl\{ \a(J) \tilde{c}^{\dag}_{J(M_1 x_1)} c_{J+\half (M_2 x_2)}
                        \nonumber  \\
        &&\qquad\qquad\qquad\qquad\qquad\qquad\qquad
             +\b(J) \tilde{d}^{\dag}_{J(M_1 x_1)} d_{J+\half (M_2 x_2)}
                        \nonumber \\
        &&\qquad\qquad\qquad\qquad\qquad\qquad\qquad
             +\gm(J) \tilde{c}^{\dag}_{J+\half (M_2 x_2)} d_{J (M_1 x_1)} \Bigr\}
                        \nonumber \\
        && + \sum_{J \geq 1}\sum_{M_1,x_1, M_2,y_2}
            \H^{\half M}_{J(M_1 x_1); J (M_2 y_2)}
             \Bigl\{ A(J) \tilde{c}^{\dag}_{J(M_1 x_1)} e_{J (M_2 y_2)}
                   \nonumber  \\
        &&\qquad\qquad\qquad\qquad\qquad\qquad\qquad
                + B(J) \tilde{e}^{\dag}_{J(M_2 y_2)} d_{J (M_1 x_1)} \Bigr\}
                     \nonumber  \\
        && + \sum_{J \geq 1}\sum_{M_1,y_1, M_2, y_2}
           \D^{\half M}_{J(M_1 y_1), J+\half (M_2 y_2)}
             C(J) \tilde{e}^{\dag}_{J(M_1 y_1)} e_{J+\half (M_2 y_2)}.
\eea
Here, $\C$, $\D$, $\E$ and $\H$ are the $SU(2)\times SU(2)$ Clebsch-Gordan coefficients, 
defined in Appendix B. The modes with the tilde are defined by:
\bba
    \tvphi_{JM} &=&\eps_M \vphi_{J-M}, 
          \nonumber \\
    \tq_{J(My)} &=& \eps_M q_{J(-My)}, 
           \nonumber  \\
    \ta_{J M} &=& \eps_M a_{J-M}, \quad \tb_{J M} = \eps_M b_{J-M}, 
          \nonumber   \\
    \tc_{J(Mx)} &=& \eps_M c_{J(-Mx)}, ~
     \td_{J(Mx)} = \eps_M d_{J(-Mx)}, ~ 
     \te_{J(My)}= \eps_M e_{J(-My)}, 
\eea
where $\eps_M=(-1)^{m-\prm}$.
The coefficients are given by:
\bba
    \a(J)&=&\sq{2J(2J+2)},
          \nonumber  \\
    \b(J)&=&-\sq{(2J+1)(2J+3)},
            \nonumber    \\
    \gm(J)&=& 1 ,
             \nonumber    \\
    A(J) &=& \sq{\fr{4J}{(2J-1)(2J+3)}},
                 \\
    B(J) &=& \sq{\fr{2(2J+2)}{(2J-1)(2J+3)}},
               \nonumber   \\
    C(J) &=& \sq{\fr{(2J-1)(2J+1)(2J+2)(2J+4)}{2J(2J+3)}}.
                \nonumber
\eea
These conformal charges satisfy the closed algebra (\ref{algebra}).

\section{Building Blocks for Scalar Fields}
\setcounter{equation}{0}
\noindent

  We first consider the scalar field sector in the physical states. 
Basic ideas on how to construct building blocks of physical states 
in a non-critical 3-brane are given here.

The standard Fock state created by acting with a creation mode on a vacuum   
is an eigenstate of the Hamiltonian. 
It belongs to a representation of the rotation group on $S^3$, $SU(2)\times SU(2)$.     
However, the charges for the special conformal transformations, $Q_M$, map 
a creation mode to another creation mode belonging to a different representation.   
Therefore, the $Q_M$ conditions in (\ref{phys-conditions}) are non-trivial, and 
we must impose them in each sector, while the $H$ and $R_{MN}$ conditions are 
imposed last after combining all sectors.

Let us seek creation operators that commute with $Q_M$. 
The commutator of $Q_M$ and the creation mode $\vphi^{\dag}_{JM}$ is  
calculated as 
\bb
   [ Q_M, \tvphi^{\dag}_{JM_1} ]
   = \sq{2J(2J+1)}\sum_{M_2} \eps_{M_1}
     \C^{\half M}_{J -M_1, J-\half M_2}\tvphi^{\dag}_{J-\half M_2}. 
\ee
Thus, only the lowest mode, $\vphi^{\dag}_{00}$, commutes with $Q_M$.

Consider creation operators with the scalar index $J N$, constructed 
from the products of two creation modes. 
The general operator with level $H=2L+2$ is given by 
\bb
     \tilde{\Phi}^{[2L+2]\dag}_{J N} 
     = \sum_{K=0}^L \sum_{M_1,M_2} \bar{f}(L,K)
        \C^{J N}_{L-K M_1, K M_2}
        \tvphi^{\dag}_{L-K M_1} \tvphi^{\dag}_{K M_2}.
            \label{Phi}
\ee
An operator without the tilde is defined by 
$\Phi^{[2L+2]}_{J N} =\eps_N \tilde{\Phi}^{[2L+2]}_{J -N}$. 
Because the property of the $SU(2) \times SU(2)$ Clebsch-Gordan coefficient, 
$\C^{JN}_{J_1M_1, J_2M_2}=\C^{J-N}_{J_1-M_1, J_2-M_2}$, and $N=M_1+M_2$, such 
that $\eps_N=\eps_{M_1}\eps_{M_2}$,  
$\Phi^{[2L+2]\dag}_{J N}$ is expressed by (\ref{Phi}) 
without the tildes on $\vphi^\dag$'s. 
The function $\bar{f}$ is defined by
\bb
    \bar{f}(L,K) = \fr{f(L,K)}{\sq{(2L-2K+1)(2K+1)}},
\ee
where $f$ satisfies the symmetric condition
\bb
        f(L,K)=f(L,L-K).
        \label{sym-f}
\ee
Because of the triangular conditions for the Clebsch-Gordan coefficient of 
type $\C$ (\ref{triangular-C}), this operator exists for $J \leq L$.

   Here, a graphical representation is introduced to help our intuitive understanding 
of the structure of the physical states. The creation and annihilation modes are described as
%%%%%%%%%%%%%%(Figure Scalar-1)%%%%%%%%%%%
\begin{center}
\begin{picture}(400,100)(0,0)
\ArrowLine(40,65)(40,35)\GCirc(40,35){3}{1} 
    \Text(40,72)[]{$J$}
\Text(80,50)[]{$=\vphi^{\dag}_{JM}$,}
\ArrowLine(130,35)(130,65)\GCirc(130,35){3}{1}
        \Text(130,72)[]{$J$}
\Text(170,50)[]{$=\tvphi^{\dag}_{JM}$,}
\ArrowLine(220,65)(220,35) 
        \Line(217,38)(223,32)\Line(223,38)(217,32)
    \Text(220,72)[]{$J$}
\Text(260,50)[]{$=\vphi_{JM}$,}
\ArrowLine(310,35)(310,65)
        \Line(307,38)(313,32)\Line(313,38)(307,32)
     \Text(310,72)[]{$J$}
\Text(350,50)[]{$=\tvphi_{JM}$.}
\end{picture} 
\end{center}
%%%%%%%%%%%%%%%%%%%%%%%%%%%%%%%%%%
The $SU(2)\times SU(2)$ Clebsch-Gordan coefficients of type $\C$ are 
%%%%%%%%%%%%%%%(Figure Scalar-2)%%%%%%%%%%%%%
\begin{center}
\begin{picture}(400,100)(0,0)
\ArrowLine(70,45)(70,75)
     \Text(70,82)[]{$J$}
  \ArrowLine(40,25)(70,45)
       \Text(40,16)[]{$J_1$} 
  \ArrowLine(100,25)(70,45)
         \Text(100,16)[l]{$J_2$} 
\Text(140,50)[]{$= \C^{JM}_{J_1M_1, J_2M_2}$,} 

\ArrowLine(250,45)(250,75)
     \Text(250,82)[]{$J$}
  \ArrowLine(220,25)(250,45)
       \Text(220,16)[]{$J_1$} 
  \ArrowLine(250,45)(280,25)
         \Text(280,16)[l]{$J_2$} 
\Text(320,50)[]{$= \eps_{M_2}\C^{JM}_{J_1M_1, J_2 -M_2}$.}
\end{picture} 
\end{center}
%%%%%%%%%%%%%%%%%%%%%%%%%%%%%%%%%%   
Using these graphs, the creation operator (\ref{Phi}) 
is expressed as
%%%%%%%%%%%%%%(Figure Scalar-3)%%%%%%%%%%%
\begin{center}
\begin{picture}(320,100)(0,0)
\Text(120,50)[]{$\tilde{\Phi}^{[2L+2]\dag}_{J N} = \sum_K \bar{f}(L,K)$} 
\ArrowLine(210,45)(210,75)
     \Text(210,82)[]{$J$}
  \ArrowLine(180,25)(210,45)\GCirc(180,25){3}{1}
       \Text(180,14)[]{$L-K$} 
  \ArrowLine(240,25)(210,45)\GCirc(240,25){3}{1}
       \Text(240,14)[l]{$K$.}
\end{picture}  
\end{center}
%%%%%%%%%%%%%%%%%%%%%%%%%%%%%%%%%%
Also, the conformal charge, $Q_M$, is expressed as
%%%%%%%%%%%%%(Figure Scalar-4)%%%%%%%%%%%
\begin{center}
\begin{picture}(300,100)(0,0)
\Text(100,50)[]{$Q_M = \sum_J \rho(J)$}
\ArrowLine(170,45)(170,75)
     \Text(170,82)[]{$\half$}
  \ArrowLine(140,25)(170,45)\GCirc(140,25){3}{1}
       \Text(140,14)[]{$J$} 
  \ArrowLine(170,45)(200,25)
           \Line(197,28)(203,22)\Line(203,28)(197,22)
         \Text(200,14)[]{$J+\half$,} 
\end{picture} 
\end{center}
%%%%%%%%%%%%%%%%%%%%%%%%%%%%%%%%%%
where $\rho(J)=\sq{(2J+1)(2J+2)}$.

  The commutator of $Q_M$ and the creation operator (\ref{Phi}) is 
calculated as 
\bba
  && [ Q_M , \tilde{\Phi}^{[2L+2]\dag}_{J N} ]
         \nonumber \\
  &&= \sum_{K=0}^L \sum_{M_1,M_2}
      \tvphi^{\dag}_{L-K-\half M_1}\tvphi^{\dag}_{K M_2}
               \nonumber \\
  && \quad \times
      \sum_S  \biggl\{ f(L,K) \sq{\fr{2L-2K}{2K+1}}
                 \eps_S \C^{\half M}_{L-K-\half M_1, L-K -S}
                            \C^{J N}_{L-K S, K M_2}
                \nonumber  \\
  && \quad
         + f \left( L, K+\half \right) \sq{\fr{2K+1}{2L-2K}}
               \eps_S \C^{\half M}_{K M_2, K+\half -S}
          \C^{J N}_{K+\half S, L-K-\half M_1}  \biggr\} . 
                \label{Q-Phi}
\eea
This equation is graphically expressed in Fig.1.
%%%%%%%%%%%%%%%(Figure 1, Scalar-5)%%%%%%%%%%%%%%
\begin{center}
\begin{picture}(400,220)(0,-10)
\Text(0,200)[l]{$[Q_M,\tilde{\Phi}^{[2L+2]\dag}_{JN} ]$}
\Text(0,150)[l]{$= \sum_K \rho(L-K-\half)\bar{f}(L,K)$}
   \ArrowLine(160,150)(140,175)\Text(140,184)[]{$\half$}
   \ArrowLine(140,125)(160,150)\GCirc(140,125){3}{1}\Text(140,114)[]{$L-K-\half$}
   \ArrowLine(160,150)(200,150)\Text(180,159)[]{$L-K$}
   \ArrowLine(200,150)(220,175)\Text(220,184)[]{$J$}
   \ArrowLine(220,125)(200,150)\GCirc(220,125){3}{1}\Text(220,114)[]{$K$}
\Text(210,150)[l]{$+\sum_K \rho(K-\half)\bar{f}(L,K)$} 
   \ArrowLine(340,150)(320,175)\Text(320,184)[]{$\half$}
   \ArrowLine(320,125)(380,150)\GCirc(320,125){3}{1}\Text(320,114)[]{$L-K$}
   \ArrowLine(340,150)(380,150)\Text(360,159)[]{$K$}
   \ArrowLine(380,150)(400,175)\Text(400,184)[]{$J$}
   \ArrowLine(400,125)(340,150)\GCirc(400,125){3}{1}\Text(400,114)[]{$K-\half$}
  
\Text(0,50)[l]{$=\sum_K \Biggl\{ \rho(L-K-\half)\bar{f}(L,K)$}
   \ArrowLine(160,50)(140,75)\Text(140,84)[]{$\half$}
   \ArrowLine(140,25)(160,50)\GCirc(140,25){3}{1}\Text(140,14)[]{$L-K-\half$}
   \ArrowLine(160,50)(200,50)\Text(180,59)[]{$L-K$}
   \ArrowLine(200,50)(220,75)\Text(220,84)[]{$J$}
   \ArrowLine(220,25)(200,50)\GCirc(220,25){3}{1}\Text(220,14)[]{$K$}
\Text(220,50)[l]{$+ \rho(K)\bar{f}(L,K+\half)$} 
   \ArrowLine(340,50)(320,75)\Text(320,84)[]{$\half$}
   \ArrowLine(320,25)(380,50)\GCirc(320,25){3}{1}\Text(320,14)[]{$L-K-\half$}
   \ArrowLine(340,50)(380,50)\Text(360,59)[]{$K+\half$}\Text(410,50)[]{$\Biggr\}$}
   \ArrowLine(380,50)(400,75)\Text(400,84)[]{$J$}
   \ArrowLine(400,25)(340,50)\GCirc(400,25){3}{1}\Text(400,14)[]{$K$}
\end{picture} 
\\
Figure 1: Commutator of the charge, $Q_M$, and the 
operator, $\tilde{\Phi}^{[2L+2]\dag}_{JN}$. 
\end{center}
%%%%%%%%%%%%%%%%%%%%%%%%%%%%%%%%%%

Let us seek a function $f$ that makes the r.h.s. of Eq.(\ref{Q-Phi}) vanish. 
To find  such a $f$, crossing relations among the $SU(2)\times SU(2)$ Clebsch-Gordan 
coefficients of type $\C$ are useful. 
Here, we consider the integral of the product of four scalar harmonics, 
\bb
    \int_{S^3} d\Om_3 Y^*_{\half M}Y_{J_1M_1}Y_{J_2M_2}Y^*_{JN}.  
\ee
Using the product expansion 
\bb
    Y^*_{\half M} Y_{J_1M_1} = \fr{1}{\sq{{\rm Vol}(S^3)}}
     \sum_{I=J_1\pm\half}\sum_S \eps_S \C^{\half M}_{J_1M_1, I-S}Y_{IS}, 
\ee
where the product is taken at the same point, and $S=(s, \prs)$, we obtain 
the crossing relation (Fig.2)
\bb
   \sum_{I=J_1\pm\half}\sum_S \eps_S \C^{\half M}_{J_1M_1, I-S}
              \C^{J N}_{IS,J_2M_2}
   =\sum_{I=J_2\pm\half}\sum_S \eps_S \C^{\half M}_{J_2M_2, I-S}
              \C^{J N}_{IS,J_1M_1}.
        \label{cross-CC}
\ee
%%%%%%%%%%%%%%%(Figure 2, Scalar-6)%%%%%%%%%%%%%%
\begin{center}
\begin{picture}(350,100)(0,0)
\Text(50,50)[l]{$\sum_I$}
   \ArrowLine(90,50)(70,75)\Text(70,82)[]{$\half$}
   \ArrowLine(70,25)(90,50)\Text(70,16)[]{$J_1$}
   \ArrowLine(90,50)(130,50)\Text(110,57)[]{$I$}
   \ArrowLine(130,50)(150,75)\Text(150,82)[]{$J$}
   \ArrowLine(150,25)(130,50)\Text(150,16)[]{$J_2$}
\Text(170,50)[l]{$= ~~\sum_I$} 
   \ArrowLine(230,50)(210,75)\Text(210,82)[]{$\half$}
   \ArrowLine(210,25)(270,50)\Text(210,16)[]{$J_1$}
   \ArrowLine(230,50)(270,50)\Text(250,57)[]{$I$}
   \ArrowLine(270,50)(290,75)\Text(290,82)[]{$J$}
   \ArrowLine(290,25)(230,50)\Text(290,16)[]{$J_2$}
\end{picture}\\
Figure 2: Crossing relation (\ref{cross-CC}).
\end{center}
%%%%%%%%%%%%%%%%%%%%%%%%%%%%%%%%%%

   Consider relation (\ref{cross-CC}) with the values $J_1=L-K-\half$ and $J_2=K$. 
In this case, the intermediate values, $I$, are $L-K-1$ and $L-K$ for the l.h.s., 
and  $K-\half$ and $K+\half$ for the r.h.s. of this equation. 
To make the commutator (\ref{Q-Phi}) vanish, crossing relations 
with the intermediate values, $L-K$ and $K+\half$, are required.       
For the general value of $J$, this condition is not satisfied. 
The crossing relation that we seek is obtained if we take $J=L$
because of the triangular conditions for the Clebsch-Gordan coefficients of 
type $\C$ (\ref{triangular-C}).

Using this crossing relation we find that if $f$ satisfies the recursion relation,
\bb
    f \left( L,K+\half \right) = -\fr{2L-2K}{2K+1} f(L,K),
          \label{recursion-f}
\ee
the commutator (\ref{Q-Phi}) vanishes. 
By solving this recursion relation, we obtain
\bb
     f(L,K)=(-1)^{2K} \left( \begin{array}{c}
                                     2L \\
                                     2K
                                     \end{array} \right)
                \label{f(L,K)}
\ee
up to the $L$-dependent normalization. 
Note that this solution satisfies the equation $f(L,L-K)=(-1)^{2L}f(L,K)$. 
However, $f$ must satisfy the symmetric condition (\ref{sym-f}). Hence,  
$f$ is given by Eq.(\ref{f(L,K)}) for integer $L$, 
while $f$ vanishes for half integer $L$.

Thus, we find that the creation operators (\ref{Phi})  
commute with $Q_M$ only when $J=L$, and $f$ is given by (\ref{f(L,K)}), 
where $L$ is a zero or positive integer.  
Hereafter, we express these operators as 
\bba
     \tilde{\Phi}^{\dag}_{L N} \equiv \tilde{\Phi}^{[2L+2]\dag}_{L N}
     &=& \sum_{K=0}^L \sum_{M_1,M_2}
               \fr{(-1)^{2K}}{\sq{(2L-2K+1)(2K+1)}}
                      \left( \begin{array}{c}
                                     2L \\
                                     2K
                                     \end{array} \right)
             \nonumber \\
     && \qquad\qquad \times
        \C^{L N}_{L-K M_1, K M_2}
        \tvphi^{\dag}_{L-K M_1} \tvphi^{\dag}_{K M_2} .
\eea
If we impose $Z_2$ symmetry, $X \leftrightarrow -X$,
the operator $(\vphi^{\dag}_{00})^n$ with  odd $n$ is excluded, while that with
even $n$ is generated from $\Phi^{\dag}_{00}=(\vphi^{\dag}_{00})^2$.
\begin{center}
\begin{tabular}{|c|c|}  \hline
rank of tensor index & $0$        \\ \hline
creation op.  & $\Phi^\dag_{LN}$ \\
level $(L \in {\bf Z}_{\geq 0})$ &  $2L+2$ \\ \hline
\end{tabular} \\ 
Table 1: Building blocks in the scalar field sector. 
\end{center}

These operators, $\Phi^\dag_{LN}$, provide building blocks of phyical states 
in the scalar field sector. 
Any state satisfying the $Q_M$-invariance condition is created by these operators.   
It is a Hamiltonian eigenstate belonging  
to a certain representation of the rotation group on $S^3$.     
The rotation invariant state is obtained by
contracting out all scalar indices, $N$, using the $SU(2)\times SU(2)$ Clebsch-Gordan 
coefficients. For example, such invariant combinations are constructed as (Fig.3)
\bb
     \sum_N \tilde{\Phi}^{\dag}_{L N} \Phi^{\dag}_{L N} |0 \rangle , \qquad
     \sum_{N_1,N_2,N_3} \C^{L_3 N_3}_{L_1 N_1, L_2 N_2}
     \Phi^{\dag}_{L_3 N_3} \tilde{\Phi}^{\dag}_{L_1 N_1}
       \tilde{\Phi}^{\dag}_{L_2 N_2}  |0 \rangle ,
             \label{Phi-states}
\ee
where $|0 \rangle$ indicates the standard Fock vacuum. 
%%%%%%%%%%%%%%(Figure 3, Scalar-7)%%%%%%%%%%%
\begin{center}
\begin{picture}(320,100)(0,0)
\CArc(50,50)(10,0,360) 
    \Text(50,50)[]{$\Phi^\dag$}
\ArrowLine(60,50)(100,50)\Text(80,40)[]{$L$}
\CArc(110,50)(10,0,360) 
    \Text(110,50)[]{$\Phi^\dag$} 
\Text(140,50)[]{$|0 \rangle$}
 
\ArrowLine(240,35)(240,65)\Text(250,55)[]{$L_3$}
\CArc(240,75)(10,0,360) 
    \Text(240,75)[]{$\Phi^\dag$}
\ArrowLine(210,35)(240,35)\Text(220,25)[]{$L_1$}
\CArc(200,35)(10,0,360) 
    \Text(200,35)[]{$\Phi^\dag$} 
\ArrowLine(270,35)(240,35)\Text(260,25)[]{$L_2$}
\CArc(280,35)(10,0,360) 
    \Text(280,35)[]{$\Phi^\dag$}
\Text(310,50)[]{$|0 \rangle$}
\end{picture}\\
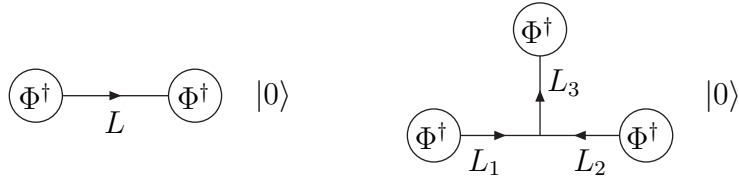
Figure 3: Examples for the $Q_M$ and $R_{MN}$ invariant states in the scalar field sector.  
The circles with $\Phi^\dag$ inside denotes the creation operators, $\Phi^\dag$.   
The arrow is defined as in the case of creation modes $\vphi^\dag$.
\end{center}
%%%%%%%%%%%%%%%%%%%%%%%%%%%%%%%%%%

In this way, an infinite number of states can be constructed from 
the building blocks, $\Phi^{\dag}_{L N}$.  
These states are graphically represented by tree diagrams in which the operators 
are connected using the Clebsch-Gordan coefficients. 
Using crossing relations, we can deform any type of tree diagram to a 
comb-type tree diagram in Fig.4. Loop diagrams also reduce to tree diagrams 
due to the properties of the Clebsch-Gordan coefficients~\cite{vmk}. 
Thus, we can deform any type of connected diagram to the comb-type tree diagram. 
Therefore, all $Q_M$-invariant states will be classified by 
the comb-type tree diagrams constructed from the building 
blocks, $\Phi^{\dag}_{L N}$, with integer $L$.

Here, we consider states restricted within the scalar field sector. In general,  
the $R_{MN}$ condition should be imposed last after all sectors are combined.
%%%%%%%%%%%%%%(Figure 4 Scalar-8)%%%%%%%%%%%
\begin{center}
\begin{picture}(320,100)(0,0)
\CArc(60,35)(10,0,360) 
    \Text(60,35)[]{$\Phi^\dag$}
\Line(70,35)(100,35)\Text(80,25)[]{$L_n$}
\Line(100,35)(100,65)\Text(115,55)[]{$L_{n-1}$}
\CArc(100,75)(10,0,360) 
    \Text(100,75)[]{$\Phi^\dag$}

\Line(100,35)(210,35)\Text(145,60)[]{$\cdot$}
                     \Text(160,60)[]{$\cdot$}
                     \Text(175,60)[]{$\cdot$}
                     \Text(190,60)[]{$\cdot$}

\Line(210,35)(210,65)\Text(220,55)[]{$L_3$}
\CArc(210,75)(10,0,360) 
    \Text(210,75)[]{$\Phi^\dag$}
\Line(210,35)(240,35) 
\Line(240,35)(240,65)\Text(250,55)[]{$L_2$}
\CArc(240,75)(10,0,360) 
    \Text(240,75)[]{$\Phi^\dag$} 
\Line(270,35)(240,35)\Text(260,25)[]{$L_1$}
\CArc(280,35)(10,0,360) 
    \Text(280,35)[]{$\Phi^\dag$}
\Text(310,50)[]{$|0 \rangle$.}
\end{picture}\\
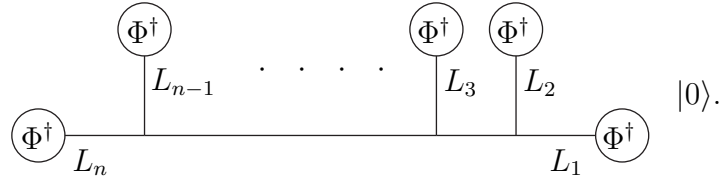
Figure 4: A state represented by a comb-type tree diagram. 
We here omit the arrow because we can turn its direction to the opposite 
using the properties of the Clebsch-Gordan coefficients, and thus it is not essential.
\end{center}
%%%%%%%%%%%%%%%%%%%%%%%%%%%%%%%%%%

\section{Building Blocks for Vector Fields}
\setcounter{equation}{0}
\noindent

Next, we consider the vector field sector. 
The commutator of the charge $Q_M$ and the creation mode $\tq^{\dag}_{J(My)}$ 
is given by
\bb
   [ Q_M, \tq^{\dag}_{J(M_1 y_1)} ]
   = - \sq{2J(2J+1)}\sum_{M_2, y_2} \eps_{M_1}
     \D^{\half M}_{J (-M_1 y_1), J-\half (M_2 y_2)}
     \tq^{\dag}_{J-\half (M_2 y_2)}.
\ee
Thus, the lowest mode, $q^{\dag}_{\half (M y)}$, is  the only creation mode 
that commutes with $Q_M$.

As in the case of the scalar field, we seek the $Q_M$-invariant creation operators 
constructed from the product of two creation modes.  
For the vector field, we must consider creation operators with 
scalar, vector and rank 2 tensor indices.  
The general forms of such creation operators with level $2L+2$ are given by 
\bba
     \tilde{\Psi}^{[2L+2]\dag}_{J N} 
     \!\!&=&\!\! -\sum_{K=\half}^{L-\half} \sum_{M_1,y_1,M_2,y_2} \bar{f}_0(L,K)
        \D^{J N}_{L-K (M_1 y_1), K (M_2 y_2)}
        \tq^{\dag}_{L-K (M_1 y_1)} \tq^{\dag}_{K (M_2 y_2)},
          \nonumber \\
          &&   \label{Psi}   \\
     {}^{\S,\A}\tilde{\Xi}^{[2L+2]\dag}_{J (Ny)} 
     \!\!&=&\!\! -\sum_{K=\half}^{L-\half} \sum_{M_1,y_1,M_2,y_2} \bar{f}_1(L,K)
        {}^{\S,\A}\V^{J (Ny)}_{L-K (M_1 y_1), K (M_2 y_2)}
        \tq^{\dag}_{L-K (M_1 y_1)} \tq^{\dag}_{K (M_2y_2)},
            \nonumber \\
         &&   \label{Xi}   \\
     \tilde{\Upsilon}^{[2L+2]\dag}_{J (Nx)} 
     \!\!&=&\!\! -\sum_{K=\half}^{L-\half} \sum_{M_1,y_1,M_2,y_2} \bar{f}_2(L,K)
        \F^{J (Nx)}_{L-K (M_1y_1), K (M_2y_2)}
        \tq^{\dag}_{L-K (M_1y_1)} \tq^{\dag}_{K (M_2y_2)}.
          \nonumber \\
       && \label{Upsilon}
\eea
These operators without the tilde are defined by the 
relations $\tilde{\Psi}^{[2L+2]}_{J N} =\eps_N \Psi^{[2L+2]}_{J -N}$, 
$\tilde{\Xi}^{[2L+2]}_{J (Ny)} =\eps_N \Xi^{[2L+2]}_{J (-Ny)}$
and $\tilde{\Upsilon}^{[2L+2]}_{J (Nx)} =\eps_N \Upsilon^{[2L+2]}_{J (-Nx)}$. 
The functions $\bar{f}_n$, with $n=0,1,2$, are defined by
\bb
    \bar{f}_n(L,K) = \fr{f_n(L,K)}{\sq{(2L-2K+1)(2K+1)}},
\ee
where $f_n$ satisfy the symmetric conditions
\bb
     f_n(L,L-K) = f_n(L,K). 
       \label{sym-fn}
\ee
The new $SU(2)\times SU(2)$ Clebsch-Gordan coefficients of  
types ${}^{\S,\A}\V$ and $\F$ are defined by
\bba
   {}^\S \V^{J (My)}_{J_1 (M_1 y_1), J_2 (M_2 y_2)} 
    &=& \sq{{\rm Vol}(S^3)} \int_{S^3} 
         \hnabla^{(i}Y^{j)*}_{J (My)} Y_{i J_1(M_1y_1)} Y_{j J_2(M_2y_2)}, 
              \\ 
   {}^\A \V^{J (My)}_{J_1 (M_1 y_1), J_2 (M_2 y_2)}
    &=& \sq{{\rm Vol}(S^3)} \int_{S^3} 
         \hnabla^{[i}Y^{j]*}_{J (My)} Y_{i J_1(M_1y_1)} Y_{j J_2(M_2y_2)}, 
              \\
   \F^{J (Mx)}_{J_1 (M_1y_1), J_2 (M_2y_2)}
    &=& \sq{{\rm Vol}(S^3)} \int_{S^3} 
          Y^{ij*}_{J (Mx)} Y_{i J_1(M_1y_1)} Y_{j J_2(M_2y_2)}. 
             \label{F}
\eea

Graphically, these operators are expressed as 
%%%%%%%%%%%%%%(Figure Vector-1-a)%%%%%%%%%%%
\begin{center}
\begin{picture}(320,100)(0,0)
\Text(120,50)[]{$\Psi^{[2L+2]\dag}_{J N} = \sum_K \bar{f}_0(L,K)$} 
\Line(210,45)(210,75)
     \Text(210,82)[]{$J$}
  \Photon(180,25)(210,45){2}{4}\GCirc(180,25){3}{1}
       \Text(180,14)[]{$L-K$} 
  \Photon(240,25)(210,45){2}{4}\GCirc(240,25){3}{1}
       \Text(240,14)[l]{$K$}
\end{picture}
\end{center}
%%%%%%%%%%%%%%%%%%%%%%%%%%%%%%%%%%
%%%%%%%%%%%%%%(Figure Vector-1-b)%%%%%%%%%%%
\begin{center}
\begin{picture}(320,100)(0,0)
\Text(120,50)[]{$\Xi^{[2L+2]\dag}_{J (Ny)} = \sum_K \bar{f}_1(L,K)$} 
\Photon(210,45)(210,75){2}{4}
     \Text(210,82)[]{$J$}
  \Photon(180,25)(210,45){2}{4}\GCirc(180,25){3}{1}
       \Text(180,14)[]{$L-K$} 
  \Photon(240,25)(210,45){2}{4}\GCirc(240,25){3}{1}
       \Text(240,14)[l]{$K$}
\end{picture} 
\end{center}
%%%%%%%%%%%%%%%%%%%%%%%%%%%%%%%%%%
%%%%%%%%%%%%%%(Figure Vector-1-c)%%%%%%%%%%%
\begin{center}
\begin{picture}(320,100)(0,0)
\Text(120,50)[]{$\Upsilon^{[2L+2]\dag}_{J (Nx)} = \sum_K \bar{f}_2(L,K)$} 
\Gluon(210,45)(210,75){3}{4}
     \Text(210,82)[]{$J$}
  \Photon(180,25)(210,45){2}{4}\GCirc(180,25){3}{1}
       \Text(180,14)[]{$L-K$} 
  \Photon(240,25)(210,45){2}{4}\GCirc(240,25){3}{1}
       \Text(240,14)[l]{$K$}
\end{picture} 
\end{center}
%%%%%%%%%%%%%%%%%%%%%%%%%%%%%%%%%%
The wavy and spiral lines indicate the vector and rank 2 tensor 
indices, respectively. The vertices denote the $SU(2)\times SU(2)$ Clebsch-Gordan 
coefficients of types, $\D$, $\V$ and $\F$, respectively. 
Each line has an arrow, but it is omitted here. It can be easily recovered.

   First, consider the creation operator with scalar index (\ref{Psi}).   
Because of the triangular conditions of type $\D$ (\ref{triangular-D}), 
this operator exists for $J \leq L$.  
The commutator of this operator and $Q_M$ is calculated as
\bba
  && [ Q_M , \tilde{\Psi}^{[2L+2]\dag}_{J N} ]
             \nonumber \\
  && = -\sum_{K=\half}^{L-\half} \sum_{M_1,y_1}\sum_{M_2,y_2}
      \tq^{\dag}_{L-K-\half (M_1 y_1)}\tq^{\dag}_{K (M_2 y_2)}
               \nonumber \\
  && \quad \times
      \sum_{V, y}  \biggl\{ f_0 (L,K) \sq{\fr{2L-2K}{2K+1}}
                 \eps_V \D^{\half M}_{L-K-\half (M_1 y_1), L-K (-V y)}
                            \D^{J N}_{L-K (V y), K (M_2 y_2)}
                \nonumber  \\
  && \quad
         + f_0 \left( L, K+\half \right) \sq{\fr{2K+1}{2L-2K}}
               \eps_V \D^{\half M}_{K (M_2 y_2), K+\half (-V y)}
          \D^{J N}_{K+\half (V y), L-K-\half (M_1 y_1)}  \biggr\} .
               \nonumber \\
  &&             \label{Q-Psi}
\eea
Therefore, according to the procedure developed in Sect.3, 
we seek a crossing relation that consists of only the type $\D$ coefficients 
in order to find a function $f_0$ that makes the r.h.s. of this commutator vanish.

Consider the integral of the product of two scalar and two vector harmonics,
\bb
     \int_{S^3} d\Om_3 Y^*_{\half M}Y^i_{J_1(M_1y_1)}Y_{i J_2(M_2y_2)}Y^*_{JN}. 
\ee
{}From the product expansion,
\bba
   && Y^*_{\half M}Y^i_{J_1(M_1y_1)}
          \nonumber \\ 
   && = -\fr{1}{\sq{{\rm Vol}(S^3)}} \sum_{I=J_1\pm\half} 
         \sum_{V,y}\eps_V \D^{\half M}_{J_1(M_1y_1), I(-Vy)} Y^i_{I(Vy)} 
            \nonumber   \\
  && \quad +\fr{1}{\sq{{\rm Vol}(S^3)}} \sum_{I=J_1} 
         \sum_S \fr{1}{2I(2I+2)} \eps_S \G^{\half M}_{J_1(M_1y_1);I-S} 
         \hnabla^i Y_{IS},
             \label{product-sv}
\eea
we obtain the crossing relation (Fig.5)
\bba
    && \sum_{I=J_1\pm\half}\sum_{V,y} \eps_V \D^{\half M}_{J_1(M_1y_1), I(-Vy)}
         \D^{J N}_{I(Vy),J_2(M_2y_2)} 
             \nonumber \\
    &&   -\sum_{I=J_1} \sum_S \fr{1}{2I(2I+2)} \eps_S \G^{\half M}_{J_1(M_1y_1);I-S}
            \G^{JN}_{J_2(M_2y_2);IS} 
             \nonumber \\
    && = [J_1(M_1y_1) \leftrightarrow J_2(M_2y_2)]. 
              \label{cross-DD}
\eea
%%%%%%%%%%%%%%%(Figure 5, Vector-2)%%%%%%%%%%%%%%
\begin{center}
\begin{picture}(350,100)(30,0)
   \Line(90,50)(70,75)\Text(70,84)[]{$\half$}
   \Photon(70,25)(90,50){2}{4}\Text(70,14)[]{$J_1$}
   \Photon(90,50)(130,50){2}{4}\Text(110,61)[]{$J_1\pm\half$}
   \Line(130,50)(150,75)\Text(150,84)[]{$J$}
   \Photon(150,25)(130,50){2}{4}\Text(150,14)[]{$J_2$}
\Text(170,50)[l]{$+$} 
   \Line(230,50)(210,75)\Text(210,84)[]{$\half$}
   \Photon(210,25)(230,50){2}{4}\Text(210,14)[]{$J_1$}
   \Line(230,50)(270,50)\Text(250,59)[]{$J_1$}
   \Line(270,50)(290,75)\Text(290,84)[]{$J$}
   \Photon(290,25)(270,50){2}{4}\Text(290,14)[]{$J_2$}
\Text(300,50)[l]{$= ~ [J_1 \leftrightarrow J_2]$}
\end{picture}\\
Figure 5: The crossing relation (\ref{cross-DD}). 
\end{center}
%%%%%%%%%%%%%%%%%%%%%%%%%%%%%%%%%%

    Substituting the values $J_1=L-K-\half$ and $J_2=K$ into this crossing relation , 
we find that the intermediate values, $I$, are restricted to 
be $L-K-1$ and $L-K$ $(K \pm \half)$ for the $\D \cdot \D$ part in the l.h.s. (r.h.s.), 
and $L-K-\half$ $(K)$ for the $\G \cdot \G$ part 
in the l.h.s. (r.h.s.), respectively.  
The crossing relation that we seek is that in which  
the $\D \cdot \D$ part in the l.h.s. (r.h.s.) has intermediate 
values of $L-K$ $(K+\half)$, and the $\G \cdot \G$ parts vanish.  
These conditions are satisfied only when $J=L$, because the triangular 
conditions of type $\D$ (\ref{triangular-D}) and type $\G$ (\ref{triangular-G})   
read 
\bba
  &&\D^{LN}_{L-K-1(Vy),K(M_2y_2)}=\D^{LN}_{K-\half(Vy),L-K-\half(M_1y_1)}=0, 
                    \\
  &&\G^{LN}_{K(M_2y_2);L-K-\half S}=\G^{LN}_{L-K-\half(M_1y_1);KS}=0.
\eea   
Using this crossing relation, we find that the commutator (\ref{Q-Psi}) 
vanishes if the function $f_0$ satisfies the same recursion relation 
to $f$ (\ref{recursion-f}) in the scalar field sector, and hence
\bb
         f_0(L,K)=f(L,K), 
\ee
where $f$ is given by Eq.(\ref{f(L,K)}), and $L$ is an integer in order to satisfy the 
symmetric condition (\ref{sym-fn}).

Thus, we obtain a creation operator with scalar index that commutes with $Q_M$ as
\bba
    \tilde{\Psi}^{\dag}_{L N} \equiv \tilde{\Psi}^{[2L+2]\dag}_{L N}     
     &=& \sum_{K=\half}^{L-\half} \sum_{M_1,y_1,M_2,y_2}
               \fr{(-1)^{2K+1}}{\sq{(2L-2K+1)(2K+1)}}
                      \left( \begin{array}{c}
                                     2L \\
                                     2K
                                     \end{array} \right)
             \nonumber \\
     && \qquad \times
        \D^{L N}_{L-K (M_1 y_1), K (M_2 y_2)}
          \tq^{\dag}_{L-K (M_1 y_1)} \tq^{\dag}_{K (M_2 y_2)},
             \label{Psi(L)}
\eea
with integer $L (\geq 2)$. Here, $L=1$ is trivial because $[Q_M, q^\dag_{\half(My)}]=0$, 
and therefore it is removed.

     In the same way, we can obtain a $Q_M$-invariant creation operator 
with  rank 2 tensor index. 
The triangular conditions of the type $\F$ coefficient (\ref{F}) can be 
obtained from the expression
\bb
        \F^{J (Mx)}_{J_1 (M_1y_1), J_2 (M_2y_2)} 
        \propto C^{J+x m}_{J_1+y_1 m_1, J_2+y_2 m_2}
                C^{J-x \prm}_{J_1-y_1 \prm_1, J_2-y_2 \prm_2},
\ee
and thus this coefficient is non-vanishing for 
\bb
         J \leq J_1+J_2,
          \label{triangular-F}
\ee
with integer $J+J_1+J_2$, where the equality is saturated only when $x$, $y_1$, $y_2$ 
have the same sign. 
Therefore, the creation operator (\ref{Upsilon}) exists for $J \leq L$.  
The commutator of this operator and $Q_M$ is calculated as 
\bba
  && [ Q_M , \tilde{\Upsilon}^{[2L+2]\dag}_{J (Nx)} ]
     = -\sum_{K=\half}^{L-\half} \sum_{M_1,y_1}\sum_{M_2,y_2}
      \tq^{\dag}_{L-K-\half (M_1 y_1)}\tq^{\dag}_{K (M_2 y_2)}
               \nonumber \\
  && \quad \times
      \sum_{V, y}  \biggl\{ f_2 (L,K) \sq{\fr{2L-2K}{2K+1}}
                 \eps_V \D^{\half M}_{L-K-\half (M_1 y_1), L-K (-V y)}
                            \F^{J (Nx)}_{L-K (V y), K (M_2 y_2)}
                \nonumber  \\
  && \qquad
         + f_2 \left( L, K+\half \right) \sq{\fr{2K+1}{2L-2K}}
               \eps_V \D^{\half M}_{K (M_2 y_2), K+\half (-V y)}
          \F^{J (Nx)}_{K+\half (V y), L-K-\half (M_1 y_1)}  \biggr\} .
                \nonumber \\
  &&        \label{Q-Upsilon}
\eea

The crossing relation that we need in this case is obtained from the integral 
\bb
    \int_{S^3} d\Om_3 Y^*_{\half M} Y^i_{J_1(M_1y_1)}
        Y^j_{J_2(M_2y_2)}Y^*_{ij J(Nx)} 
\ee
as (Fig.6)
\bba
    && \sum_{I=J_1\pm\half}\sum_{V,y} \eps_V 
         \D^{\half M}_{J_1(M_1y_1), I(-Vy)}
           \F^{J (Nx)}_{I(Vy),J_2(M_2y_2)} 
             \nonumber \\
    &&   +\sum_{I=J_1} \sum_S \fr{1}{2I(2I+2)} \G^{\half M}_{J_1(M_1y_1);IS}
            \eps_N \H^{IS}_{J(-Nx);J_2(M_2y_2)} 
             \nonumber \\
    && = [J_1(M_1y_1) \leftrightarrow J_2(M_2y_2)], 
                \label{cross-DF}
\eea
where the product expansion (\ref{product-sv}) is used. 
%%%%%%%%%%%%%%%(Figure 6, Vector-3)%%%%%%%%%%%%%%
\begin{center}
\begin{picture}(350,100)(30,0)
   \Line(90,50)(70,75)\Text(70,84)[]{$\half$}
   \Photon(70,25)(90,50){2}{4}\Text(70,14)[]{$J_1$}
   \Photon(90,50)(130,50){2}{4}\Text(110,61)[]{$J_1\pm\half$}
   \Gluon(130,50)(150,75){3}{4}\Text(150,84)[]{$J$}
   \Photon(150,25)(130,50){2}{4}\Text(150,14)[]{$J_2$}
\Text(170,50)[l]{$+$} 
   \Line(230,50)(210,75)\Text(210,84)[]{$\half$}
   \Photon(210,25)(230,50){2}{4}\Text(210,14)[]{$J_1$}
   \Line(230,50)(270,50)\Text(250,59)[]{$J_1$}
   \Gluon(270,50)(290,75){3}{4}\Text(290,84)[]{$J$}
   \Photon(290,25)(270,50){2}{4}\Text(290,14)[]{$J_2$}
\Text(300,50)[l]{$= ~ [J_1 \leftrightarrow J_2]$}
\end{picture}\\
Figure 6: Crossing relation (\ref{cross-DF}).
\end{center}
%%%%%%%%%%%%%%%%%%%%%%%%%%%%%%%%%%

The necessary condition for the commutator (\ref{Q-Upsilon}) to vanish is now that there is a 
crossing relation with $J_1=L-K-\half$ and $J_2=K$ that consists of only the $\D \cdot \F$ part   
with the intermediate value $I=L-K$ $(K+\half)$ in the l.h.s. (r.h.s.) of (\ref{cross-DF}). 
If we take $J=L$, we obtain the required relation, because 
the triangular conditions of type $\F$ (\ref{triangular-F}) 
and type $\H$ (\ref{triangular-H}) read 
\bba
  &&\F^{L(Nx)}_{L-K-1(Vy),K(M_2y_2)}=\F^{L(Nx)}_{K-\half(Vy),L-K-\half(M_1y_1)}=0, 
                    \\
  &&\H^{L-K-\half S}_{L(Nx);K(M_2y_2)}=\H^{KS}_{L(Nx);L-K-\half(M_1y_1)}=0.
\eea
Using this crossing relation, we find that $f_2$ must also satisfy the same recursion 
relation to $f$ in order that the commutator (\ref{Q-Upsilon}) vanishes. 
Thus, we obtain
\bb
         f_2 (L,K)=f(L,K),
\ee
where $L$ is an integer in order to satisfy the symmetric condition (\ref{sym-fn}).

Thus, the $Q_M$-invariant creation operator with rank $2$ tensor index is 
given by
\bba
     \tilde{\Upsilon}^{\dag}_{L (Nx)} \equiv \tilde{\Upsilon}^{[2L+2]\dag}_{L (N x)}
     &=& \sum_{K=\half}^{L-\half} \sum_{M_1,y_1}\sum_{M_2,y_2}
               \fr{(-1)^{2K+1}}{\sq{(2L-2K+1)(2K+1)}}
                      \left( \begin{array}{c}
                                     2L \\
                                     2K
                             \end{array} \right)
             \nonumber \\
     && \qquad \times
        \F^{L (Nx)}_{L-K (M_1 y_1), K (M_2 y_2)}
         \tq^{\dag}_{L-K (M_1 y_1)} \tq^{\dag}_{K (M_2 y_2)},
              \label{Upsilon(L)}
\eea
with integer $L (\geq 2)$.

Next, consider the creation operators with vector index (\ref{Xi}). 
The commutators of $Q_M$ and these operator  
are given by expression (\ref{Q-Psi}) with the quantities $f_0$, $\D^J_{L-K, K}$ 
and $\D^J_{K+\half, L-K-\half}$ replaced by $f_1$, ${}^{\S,\A}\V^J_{L-K, K}$ 
and ${}^{\S,\A}\V^J_{K+\half, L-K-\half}$, respectively. 
Thus, the necessary conditions for these commutators to vanish are 
that there are crossing relations that consist of only the $\D \cdot {}^{\S,\A}\V$ parts.

    Consider the following integrals: 
\bba
   && \int_{S^3} d\Om_3 Y^*_{\half M} Y^i_{J_1(M_1y_1)}
        Y^j_{J_2(M_2y_2)}\hnabla_{(i}Y^*_{j)J(My)}, 
                  \\
   && \int_{S^3} d\Om_3 Y^*_{\half M} Y^i_{J_1(M_1y_1)}
        Y^j_{J_2(M_2y_2)}\hnabla_{[i}Y^*_{j]J(My)}.
\eea
These integrals give the crossing relations (Fig.7)
\bba
    && \sum_{I=J_1\pm\half}\sum_{V,y^\pp} \eps_V \D^{\half M}_{J_1(M_1y_1), I(-Vy^\pp)}
         {}^{\S,\A}\V^{J (Ny)}_{I(Vy^\pp),J_2(M_2y_2)} 
             \nonumber \\
    &&   -\sum_{I=J_1t} \sum_S \fr{1}{2I(2I+2)} \eps_S \G^{\half M}_{J_1(M_1y_1);I-S}
             {}^{\S,\A}\U^{J(Ny)}_{J_2(M_2y_2);IS} 
             \nonumber \\
    && = [J_1(M_1y_1) \leftrightarrow J_2(M_2y_2)], 
             \label{cross-DV}
\eea
where
\bba
    {}^\S\U^{J(My)}_{J_1(M_1y_1);J_2M_2} 
    &=& \sq{{\rm Vol}(S^3)} \int_{S^3} d\Om_3 
        \hnabla^{(i}Y^{j)*}_{J(My)} Y_{i J_1(M_1y_1)} \hnabla_j Y_{J_2M_2}, 
                     \\
    {}^\A\U^{J(My)}_{J_1(M_1y_1);J_2M_2} 
    &=& \sq{{\rm Vol}(S^3)} \int_{S^3} d\Om_3 
        \hnabla^{[i}Y^{j]*}_{J(My)} Y_{i J_1(M_1y_1)} \hnabla_j Y_{J_2M_2}. 
\eea
%%%%%%%%%%%%%%%(Figure 7, Vector-4)%%%%%%%%%%%%%%
\begin{center}
\begin{picture}(350,100)(30,0)
   \Line(90,50)(70,75)\Text(70,84)[]{$\half$}
   \Photon(70,25)(90,50){2}{4}\Text(70,14)[]{$J_1$}
   \Photon(90,50)(130,50){2}{4}\Text(110,61)[]{$J_1\pm\half$}
                               \Text(125,40)[]{${}^{\S,\A}$}
   \Photon(130,50)(150,75){2}{4}\Text(150,84)[]{$J$}
   \Photon(150,25)(130,50){2}{4}\Text(150,14)[]{$J_2$}
\Text(170,50)[l]{$+$} 
   \Line(230,50)(210,75)\Text(210,84)[]{$\half$}
   \Photon(210,25)(230,50){2}{4}\Text(210,14)[]{$J_1$}
   \Line(230,50)(270,50)\Text(250,59)[]{$J_1$}
                        \Text(265,40)[]{${}^{\S,\A}$}
   \Photon(270,50)(290,75){2}{4}\Text(290,84)[]{$J$}
   \Photon(290,25)(270,50){2}{4}\Text(290,14)[]{$J_2$}
\Text(300,50)[l]{$= ~ [J_1 \leftrightarrow J_2]$}
\end{picture}\\
Figure 7: Crossing relations (\ref{cross-DV}). 
\end{center}
%%%%%%%%%%%%%%%%%%%%%%%%%%%%%%%%%%

Here, the triangular conditions for the Clebsch-Gordan coefficients 
of types ${}^{\S,\A}\V$ and  ${}^{\S,\A}\U$ are obtained from 
expression (\ref{CG-general}).  From the expression 
\bb
  {}^{\S,\A}\V^{J(My)}_{J_1(M_1y_1),J_2(M_2y_2)} 
    \propto C^{J+y m}_{J_1+y_1 m_1,J_2+y_2 m_2} 
           C^{J-y \prm}_{J_1-y_1 \prm_1,J_2-y_2 \prm_2},
\ee
this coefficient is non-vanishing for 
\bb
         J \leq J_1+J_2-\half,
         \label{triangular-V}
\ee
with half integer $J+J_1+J_2$. Also, from the expression
\bb
  {}^{\S,\A}\U^{J(My)}_{J_1(M_1y_1);J_2M_2} 
    \propto C^{J+y m}_{J_1+y_1 m_1,J_2 m_2} 
            C^{J-y \prm}_{J_1-y_1 \prm_1,J_2 \prm_2}, 
\ee
we obtain the non-vanishing condtion for this coefficient as 
\bb
      J \leq J_1+J_2,
        \label{triangular-U}
\ee
with integer $J+J_1+J_2$. The equality in (\ref{triangular-U}) is saturated only at $y=y_1$.

In order that the $Q_M$ invariant operator of type (\ref{Xi}) exists,
it is required that there is a crossing relation  
with $J_1=L-K-\half$ and $J_2=K$ that consists of only the $\D \cdot {}^{\S,\A}\V$ parts 
with the intermediate values $L-K$ and $K+\half$ in the l.h.s. and r.h.s. 
of (\ref{cross-DV}), respectively. 
The problem of finding such a crossing relation is to find a quantity $J$ satisfying 
the following conditions:
\bb
    {}^{\S,\A}\V^{J(Ny)}_{L-K(Vy^\pp),K(M_2y_2)} \neq 0, \qquad
    {}^{\S,\A}\V^{J(Ny)}_{L-K-1(Vy^\pp),K(M_2y_2)}=0,
            \label{condition-V}
\ee
and 
\bb
    {}^{\S,\A}\U^{J(Ny)}_{K(M_2y_2);L-K-\half S}=0.
             \label{condition-U}
\ee
{}From inequality (\ref{triangular-V}), 
the conditions (\ref{condition-V}) are satisfied only when $J=L-\half$. 
However, because of the triangular condition for ${}^{\S,\A}\U$ (\ref{triangular-U}), 
this value does not satisfy the second condition (\ref{condition-U}).  
Thus, we cannot make the $\G \cdot \U$ parts in the relation (\ref{cross-DV}) vanish. 
Consequently, we find that there is no $Q_M$-invariant creation operator with vector 
index of type (\ref{Xi}). 
Therefore, the only building block with vector index is given by  
the lowest vector mode, $q^\dag_{\half (Ny)}$.

Finaly, we briefly show that there is no other $Q_M$-invariant creation operators with 
tensor index of higher rank, $n$. 
Such a operator is obtained by replacing the Clebsch-Gordan coefficients of type $\D$ 
in expression (\ref{Psi}) with the genralized one, ${}^n\oD$, defined in Appendix C.
As in the previous argument, the necessary condition for the commutator of this operator 
and $Q_M$ to vanish is that there is a crossing relation that consists of 
the only $\D \cdot {}^n\oD$ parts. 
The crossing relation with such $\D \cdot {}^n\oD$ terms 
is given by the type III relation (\ref{cross-III}) derived in Appendix E. 
However, for $n \geq 3$, because of the triangular conditions of 
type ${}^n\oH$ (\ref{triangular-oH}) and type ${}^n\oG$ (\ref{triangular-oG}), 
this relation does not close  only within  the $\D \cdot {}^n\oD$ parts. 
Thus, there is no $Q_M$-invariant creation operator with index of rank $n \geq 3$.

The two types of creation operators, $\Psi^\dag_{LN}$ (\ref{Psi(L)}) 
and $\Upsilon^\dag_{L(Nx)}$ (\ref{Upsilon(L)}), and the lowest creation
mode, $q_{\half(Ny)}^\dag$, provide the building blocks of physical states 
in the vector field sector. They are summarized in Table 2.
Any $Q_M$-invariant Hamiltonian eigenstate belonging to a certain representation 
of $R_{MN}$ will be constructed from these operators,  
using the $SU(2)\times SU(2)$ Clebsch-Gordan coefficients.   
As in the case of the scalar field sector, such a state will be classified  
by the comb-type tree diagram.  
\begin{center}
\begin{tabular}{|c|ccc|}  \hline
rank of tensor index & $0$       & $1$                   & $2$         \\ \hline
creation op.  & $\Psi^\dag_{LN}$ & $q^\dag_{\half (Ny)}$ & $\Upsilon^\dag_{L(Nx)}$ \\
level $(L \in {\bf Z}_{\geq 2})$ &  $2L+2$    & $2$    & $2L+2$  \\ \hline
\end{tabular} \\ 
Table 2: Building blocks in the vector field sector. 
\end{center}

\section{Building Blocks for Gravitational Traceless Fields}
\setcounter{equation}{0}
\noindent

In this section, we construct and classify building blocks for the traceless mode sector.
The commutators of $Q_M$ and the traceless modes are given by: 
\bba
   && [Q_M, \tc^{\dag}_{J(M_1x_1)}]
         = \a\Bigl(J-\half\Bigr) \sum_{M_2,x_2} \eps_{M_1}
           \E^{\half M}_{J(-M_1x_1),J-\half (M_2x_2)}
           \tc^{\dag}_{J-\half (M_2x_2)} ,
                \\
   && [Q_M, \td^{\dag}_{J(M_1x_1)}]
          = -\gm (J) \sum_{M_2,x_2} \eps_{M_1}
           \E^{\half M}_{J(-M_1x_1),J+\half (M_2x_2)}
           \tc^{\dag}_{J+\half (M_2x_2)}
                \nonumber  \\
   && \qquad\qquad\qquad\quad
          -\b\Bigl(J-\half\Bigr) \sum_{M_2,x_2} \eps_{M_1}
           \E^{\half M}_{J(-M_1x_1),J-\half (M_2x_2)}
           \td^{\dag}_{J-\half (M_2x_2)}
                \nonumber \\
   && \qquad\qquad\qquad\quad
          -B(J)\sum_{M_2,y_2} \eps_{M_1}
           \H^{\half M}_{J(-M_1x_1);J (M_2y_2)}
           \te^{\dag}_{J (M_2 y_2)} ,
                 \\
   && [Q_M, \te^{\dag}_{J(M_1y_1)}]
          = -A(J) \sum_{M_2,x_2} \eps_{M_1}
           \H^{\half M}_{J(M_2x_2);J (-M_1y_1)}
           \tc^{\dag}_{J (M_2 x_2)}
                \nonumber  \\
   && \qquad\qquad\qquad
          -C\Bigl(J-\half\Bigr) \sum_{M_2,y_2} \eps_{M_1}
           \D^{\half M}_{J(-M_1y_1),J-\half (M_2y_2)}
           \te^{\dag}_{J-\half (M_2 y_2)}.
\eea
The only $Q_M$-invariant mode is the lowest rank 2 tensor creation mode with 
a positive metric, $c^\dag_{1(Mx)}$. No negative-metric creation 
modes, $d^\dag$ and $e^\dag$, commute with $Q_M$.

   Let us consider creation operators with tensor index $(N\veps_n)$ of rank $n$.  
We here need operators with index up to rank $4$. 
The general form of such a creation operator with level $H=2L$ constructed 
from the products of two creation modes is given by
\bba
  && \tilde{O}^{[2L]\dag}_{J (N\veps_n)}
   = \sum_{K=1}^{L-1} \sum_{M_1,x_1}\sum_{M_2,x_2} \bx_n(L,K)
      ~{}^n\oE^{J(N\veps_n)}_{L-K (M_1x_1), K(M_2,x_2)}
      \tc^{\dag}_{L-K(M_1x_1)} \tc^{\dag}_{K(M_2x_2)}
               \nonumber \\
  &&\qquad
     + \sum_{K=1}^{L-2} \sum_{M_1,x_1}\sum_{M_2,x_2} \by_n(L,K)
      ~{}^n\oE^{J(N\veps_n)}_{L-K-1 (M_1x_1), K(M_2,x_2)}
      \td^{\dag}_{L-K-1 (M_1x_1)} \tc^{\dag}_{K(M_2x_2)}
               \nonumber  \\
  &&\qquad
     + \sum_{K=1}^{L-3} \sum_{M_1,x_1}\sum_{M_2,x_2} \bz_n(L,K)
      ~{}^n\oE^{J(N\veps_n)}_{L-K-2 (M_1x_1), K(M_2,x_2)}
      \td^{\dag}_{L-K-2 (M_1x_1)} \td^{\dag}_{K(M_2x_2)}
               \nonumber  \\
  &&\qquad
     +\sum_{K=1}^{L-\fr{3}{2}} \sum_{M_1,x_1}\sum_{M_2,y_2} \bw_n(L,K)
      ~{}^n\oH^{J(N\veps_n)}_{L-K-\half (M_1x_1); K(M_2,y_2)}
      \tc^{\dag}_{L-K-\half (M_1x_1)} \te^{\dag}_{K(M_2y_2)}
               \nonumber \\
  &&\qquad
     +\sum_{K=1}^{L-\fr{5}{2}} \sum_{M_1,x_1}\sum_{M_2,y_2} \bu_n(L,K)
      ~{}^n\oH^{J(N\veps_n)}_{L-K-\fr{3}{2} (M_1x_1); K(M_2,y_2)}
      \td^{\dag}_{L-K-\fr{3}{2} (M_1x_1)} \te^{\dag}_{K(M_2y_2)}
               \nonumber \\
  &&\qquad
     +\sum_{K=1}^{L-2} \sum_{M_1,y_1}\sum_{M_2,y_2} \bv_n(L,K)
      ~{}^n\oD^{J(N\veps_n)}_{L-K-1 (M_1y_1), K(M_2,y_2)}
      \te^{\dag}_{L-K-1(M_1y_1)} \te^{\dag}_{K(M_2y_2)} ,
              \nonumber \\
  &&    \label{O}
\eea
where new Clebsch-Gordan coefficients, ${}^n\oE$, ${}^n\oH$, ${}^n\oD$, 
are defined in Appendix C, and also their non-vanishing conditions used 
in this section are summarized there.
The unknown functions are defined as 
\bba
      \bx_n(L,K) &=& \fr{x_n(L,K)}{\sq{(2L-2K+1)(2K+1)}},
           \label{bx}   \\
      \by_n(L,K) &=& \fr{y_n(L,K)}{\sq{(2L-2K-1)(2K+1)}},
           \label{by}  \\
      \bz_n(L,K) &=& \fr{z_n(L,K)}{\sq{(2L-2K-3)(2K+1)}},
                 \\
      \bw_n(L,K) &=& \fr{w_n(L,K)}{\sq{(2L-2K)(2K+1)}},
           \label{bw}   \\
      \bu_n(L,K) &=& \fr{u_n(L,K)}{\sq{(2L-2K-2)(2K+1)}},
              \\
      \bv_n(L,K) &=& \fr{v_n(L,K)}{\sq{(2L-2K-1)(2K+1)}}.
           \label{bv}
\eea
Here, $x_n$, $z_n$ and $v_n$ satisfy the following symmetric conditions: 
\bba
   x_n(L,K)&=&x_n(L,L-K), 
        \label{sym-x}  \\ 
   z_n(L,K)&=&z_n(L,L-K-2), 
        \label{sym-z} \\
   v_n(L,K)&=&v_n(L,L-K-1).
        \label{sym-v}
\eea

  The commutator of $Q_M$ and this creation operator is computed as
\bba
  && [ Q_M, \tilde{O}^{[2L]\dag}_{J (N\veps_n)} ]
          \nonumber \\
  && = \sum_{K=1}^{L-\fr{3}{2}} \sum_{M_1,x_1}\sum_{M_2,x_2}
          \tc^\dag_{L-K-\half (M_1x_1)}\tc^\dag_{K(M_2x_2)}
           \nonumber  \\
  && \quad \times \biggl[ 
       2 \bx_n(L,K)\a\left( L-K-\half \right) \sum_{T,x} \eps_T 
       \E^{\half M}_{L-K-\half(M_1x_1),L-K(-Tx)}
           \nonumber \\
  && \qquad\qquad\qquad\qquad\qquad\qquad\qquad\quad \times     
       {}^n\oE^{J(N\veps_n)}_{L-K(Tx),K(M_2x_2)}
           \nonumber \\
  &&\qquad  -\by_n(L,K)\gm(L-K-1) \sum_{T,x} \eps_T
       \E^{\half M}_{L-K-\half(M_1x_1),L-K-1(-Tx)}
           \nonumber \\
  && \qquad\qquad\qquad\qquad\qquad\qquad\qquad \times
       {}^n\oE^{J(N\veps_n)}_{L-K-1(Tx),K(M_2x_2)}
           \nonumber \\
  &&\qquad  -\bw_n(L,K) A(K) \sum_{V,y} \eps_V
       \H^{\half M}_{K(M_2x_2);K(-Vy)}
       {}^n\oH^{J(N\veps_n)}_{L-K-\half(M_1x_1);K(Vy)} \biggr]
           \nonumber \\
  && + \sum_{K=1}^{L-\fr{7}{2}} \sum_{M_1,x_1}\sum_{M_2,x_2}
          \td^\dag_{L-K-\fr{5}{2} (M_1x_1)}\td^\dag_{K(M_2x_2)}
           \nonumber  \\
  && \quad \times \biggl[ 
       -2 \bz_n(L,K)\b\left( L-K-\fr{5}{2} \right) \sum_{T,x} \eps_T 
       \E^{\half M}_{L-K-\fr{5}{2}(M_1x_1),L-K-2(-Tx)}
           \nonumber \\
  && \qquad\qquad\qquad\qquad\qquad\qquad\qquad\qquad \times
       {}^n\oE^{J(N\veps_n)}_{L-K-2(Tx),K(M_2x_2)} \biggr]
           \nonumber \\
  && + \sum_{K=1}^{L-\fr{5}{2}} \sum_{M_1,y_1}\sum_{M_2,y_2}
          \te^\dag_{L-K-\fr{3}{2} (M_1y_1)}\te^\dag_{K(M_2y_2)}
           \nonumber  \\
  && \quad \times \biggl[
        -\bu_n(L,K) B \left( L-K-\fr{3}{2} \right) \sum_{T,x} \eps_T
        \H^{\half M}_{L-K-\fr{3}{2}(-Tx);L-K-\fr{3}{2}(M_1y_1)}
           \nonumber \\
  && \qquad\qquad\qquad\qquad\qquad\qquad\qquad\qquad \times
        {}^n\oH^{J(N\veps_n)}_{L-K-\fr{3}{2}(Tx);K(M_2y_2)}
           \nonumber \\
  &&\qquad -2\bv_n(L,K) C\left( L-K-\fr{3}{2} \right) \sum_{V,y} \eps_V
        \D^{\half M}_{L-K-\fr{3}{2}(M_1y_1),L-K-1(-Vy)}
           \nonumber \\
  && \qquad\qquad\qquad\qquad\qquad\qquad\qquad\qquad \times        
        {}^n\oD^{J(N\veps_n)}_{L-K-1(Vy),K(M_2y_2)} \biggr]
          \nonumber \\
  && + \sum_{K=1}^{L-\fr{5}{2}} \sum_{M_1,x_1}\sum_{M_2,x_2}
          \td^\dag_{L-K-\fr{3}{2} (M_1x_1)}\tc^\dag_{K(M_2x_2)}
           \nonumber  \\
  && \quad \times \biggl[ 
       -\by_n(L,K) \b\left( L-K-\fr{3}{2} \right) \sum_{T,x} \eps_T 
       \E^{\half M}_{L-K-\fr{3}{2}(M_1x_1),L-K-1(-Tx)}
           \nonumber \\
  && \qquad\qquad\qquad\qquad\qquad\qquad\qquad\qquad \times
       {}^n\oE^{J(N\veps_n)}_{L-K-1(Tx),K(M_2x_2)}
           \nonumber \\
  &&\qquad  +\by_n \left( L,K+\half \right) \a(K) \sum_{T,x} \eps_T
       \E^{\half M}_{K(M_2x_2),K+\half (-Tx)}
           \nonumber \\
  && \qquad\qquad\qquad\qquad\qquad\qquad\qquad \times       
       {}^n\oE^{J(N\veps_n)}_{K+\half (Tx),L-K-\fr{3}{2}(M_1x_1)}
           \nonumber \\
  &&\qquad  -2\bz_n \left( L,K-\half \right) \gm\left( K-\half \right) \sum_{T,x} \eps_T
       \E^{\half M}_{K(M_2x_2),K-\half (-Tx)}
           \nonumber \\
  && \qquad\qquad\qquad\qquad\qquad\qquad\qquad\qquad \times       
       {}^n\oE^{J(N\veps_n)}_{K-\half (Tx),L-K-\fr{3}{2}(M_1x_1)}
           \nonumber \\
  &&\qquad  -\bu_n(L,K) A(K) \sum_{V,y} \eps_V
       \H^{\half M}_{K(M_2x_2);K(-Vy)}
       {}^n\oH^{J(N\veps_n)}_{L-K-\fr{3}{2}(M_1x_1);K(Vy)} \biggr]
           \nonumber \\
  &&+\sum_{K=1}^{L-2} \sum_{M_1,x_1}\sum_{M_2,y_2}
        \tc^\dag_{L-K-1(M_1x_1)}\te^\dag_{K(M_2y_2)}
           \nonumber \\
  &&\quad \times \biggl[
       -\by_n(L,L-K-1) B(K) \sum_{T,x} \eps_T
       \H^{\half M}_{K(-Tx);K(M_2y_2)}
           \nonumber \\
  && \qquad\qquad\qquad\qquad\qquad\qquad\qquad\quad \times
       {}^n\oE^{J(N\veps_n)}_{K(Tx),L-K-1(M_1x_1)}
          \nonumber \\
  &&\qquad  +\bw_n(L,K) \a(L-K-1) \sum_{T,x} \eps_T
       \E^{\half M}_{L-K-1(M_1x_1),L-K-\half(-Tx)}
           \nonumber \\
  && \qquad\qquad\qquad\qquad\qquad\qquad\qquad\quad \times       
       {}^n\oH^{J(N\veps_n)}_{L-K-\half(Tx);K(M_2y_2)}
           \nonumber \\
  &&\qquad  -\bw_n \left( L,K+\half \right) C(K) \sum_{V,y} \eps_V
       \D^{\half M}_{K(M_2y_2),K+\half(-Vy)}
           \nonumber \\
  && \qquad\qquad\qquad\qquad\qquad\qquad\qquad \times
       {}^n\oH^{J(N\veps_n)}_{L-K-1(M_1x_1);K+\half(Vy)}
           \nonumber \\
  &&\qquad  -\bu_n(L,K) \gm\left( L-K-\fr{3}{2} \right) \sum_{T,x} \eps_T
       \E^{\half M}_{L-K-1(M_1x_1),L-K-\fr{3}{2}(-Tx)}
           \nonumber \\
  && \qquad\qquad\qquad\qquad\qquad\qquad\qquad\quad \times       
       {}^n\oH^{J(N\veps_n)}_{L-K-\fr{3}{2}(Tx);K(M_2y_2)}
           \nonumber \\
  &&\qquad  -2\bv_n(L,K) A(L-K-1) \sum_{V,y} \eps_V
       \H^{\half M}_{L-K-1(M_1x_1);L-K-1(-Vy)}
           \nonumber \\
  && \qquad\qquad\qquad\qquad\qquad\qquad\qquad\quad \times       
       {}^n\oD^{J(N\veps_n)}_{L-K-1(Vy),K(M_2y_2)} \biggr]
           \nonumber \\
  &&+\sum_{K=1}^{L-3} \sum_{M_1,x_1}\sum_{M_2,y_2}
        \td^\dag_{L-K-2(M_1x_1)}\te^\dag_{K(M_2y_2)}
           \nonumber \\
  &&\quad \times \biggl[
       -2\bz_n(L,K) B(K) \sum_{T,x} \eps_T
       \H^{\half M}_{K(-Tx);K(M_2y_2)}
       {}^n\oE^{J(N\veps_n)}_{K(Tx),L-K-2(M_1x_1)}
          \nonumber \\
  &&\qquad  -\bu_n(L,K) \b(L-K-2) \sum_{T,x} \eps_T
       \E^{\half M}_{L-K-2(M_1x_1),L-K-\fr{3}{2}(-Tx)}
           \nonumber \\
  && \qquad\qquad\qquad\qquad\qquad\qquad\qquad\quad \times       
       {}^n\oH^{J(N\veps_n)}_{L-K-\fr{3}{2}(Tx);K(M_2y_2)}
           \nonumber \\
  &&\qquad  -\bu_n \left( L,K+\half \right) C(K) \sum_{V,y} \eps_V
       \D^{\half M}_{K(M_2y_2),K+\half(-Vy)}
           \nonumber \\
  && \qquad\qquad\qquad\qquad\qquad\qquad\qquad \times       
       {}^n\oH^{J(N\veps_n)}_{L-K-2(M_1x_1);K+\half(Vy)} \biggr].   
           \label{Q-O}
\eea

In the following,  for each $n$, we seek the functions ($x_n$, $y_n$, $z_n$, $w_n$, $u_n$ 
and $v_n$) that make this commutator vanish.

\paragraph{Building blocks with scalar index $(n=0)$:}
Let us first consider the creation operator with scalar index. 
Because of the triangular conditions for the $SU(2)\times SU(2)$ Clebsch-Gordan 
coefficients, this operator vanishes for $J > L$. 
If we take $J=L$, the only terms with the function $\bx_0$ 
in the operator (\ref{O}) survives, because  
\bba
    && {}^0\oE^{LN}_{L-K-1(M_1x_1),K(M_2x_2)}
        ={}^0\oE^{LN}_{L-K-2(M_1x_1),K(M_2x_2)}=0, 
                         \\
    && {}^0\oH^{LN}_{L-K-\half (M_1x_1);K(M_2y_2)}
         ={}^0\oH^{LN}_{L-K-\fr{3}{2}(M_1x_1);K(M_2y_2)}=0,
                         \\
    && {}^0\oD^{LN}_{L-K-1(M_1y_1),K(M_2y_2)}=0.
\eea
Thus, the only non-vanishing component in the commutator (\ref{Q-O}) is the first term 
in the $c^\dag c^\dag$ part.

To find a function $x_0$ that make this term vanish, 
the crossing relation of type I (\ref{cross-I}) for $n=0$
derived in the Appendix E is used. 
Consider the case of $J_1=L-K-\half$ and $J_2=K$ in this relation. 
If we take $J=L$, we obtain the crossing relation that consists of  
only the $\E \cdot {}^0\oE$ part with the intermediate value $I=L-K$ $(K+\half)$ 
for the l.h.s (r.h.s.). 
Using this crossing relation and the symmetric condition (\ref{sym-x}), 
we find that the commutator vanishes if the function 
satisfies the recursion relation,
\bb
   x_0 \left( L,K+\half \right) 
   = -\sq{ \fr{(2L-2K-1)(2L-2K)}{2K(2K+1)} } x_0(L,K).
            \label{recursion-x}
\ee 
Solving this equation, we obtain 
\bb
   x_0(L,K)=x(L,K)
\ee
with 
\bb
      x(L,K)= (-1)^{2K}\sq{ \left( \begin{array}{c}
                                     2L \\
                                     2K
                                     \end{array} \right)
                             \left(   \begin{array}{c}
                                     2L-2 \\
                                     2K-1
                                     \end{array} \right) }, 
              \label{x}
\ee
up to the $L$-dependent normalization. 
To satisfy the symmetric condition (\ref{sym-x}), $L$ must be an integer. 
For a half-integer $L$, this function vanishes.

Thus, we obtain the $Q_M$-invariant creation operator with scalar index, 
denoted by $\tilde{A}^\dag_{L N}$, as 
\bb
   \tilde{A}^{\dag}_{L N}
   = \sum_{K=1}^{L-1} \sum_{M_1,x_1}\sum_{M_2,x_2} \bx(L,K)
      \E^{LN}_{L-K (M_1x_1), K(M_2,x_2)}
      \tc^{\dag}_{L-K(M_1x_1)} \tc^{\dag}_{K(M_2x_2)},
\ee 
with integer $L (\geq 3)$. Here, $L=2$ is trivial because $[Q_M, c^\dag_{1(Mx)}]=0$, 
and therefore it is removed. The function $\bx$ is defined as in (\ref{bx}), 
and the definition ${}^0\oE =\E$ is used.

   Next, we consider the case of $J=L-1$. Because 
\bba
    && {}^0\oE^{L-1 N}_{L-K-2(M_1x_1),K(M_2x_2)}=0, \\ 
    && {}^0\oH^{L-1 N}_{L-K-\fr{3}{2}(M_1x_1);K(M_2y_2)}=0, 
\eea
the terms with the functions $\bz_0$ and $\bu_0$ vanish.  
Therefore, the $d^\dag d^\dag$ and $d^\dag e^\dag$ terms in the commutator (\ref{Q-O}) 
trivially vanish, and four terms with the functions $\bx_0$, $\by_0$, $\bw_0$ and $\bv_0$, 
i.e. $c^\dag c^\dag$, $e^\dag e^\dag$, $d^\dag c^\dag$ and $c^\dag e^\dag$, survive. 
We first consider the $c^\dag c^\dag$ term. 
To find a solution for which this term vanishes, 
we use the crossing relation of type I (\ref{cross-I}) for $n=0$ with the 
values $J=L-1$, $J_1=L-K-\half$ and $J_2=K$, and  
the symmetric conditions of $x_0$ (\ref{sym-x}). 
We then find that this term vanishes 
when $x_0$ satisfies the same recursion relation to Eq.(\ref{recursion-x}),
so that $x_0(L,K)=x(L,K)$, and $y_0$ and $w_0$ satisfy the equations,
\bba
   y_0(L,K) &=& y(L,K), \\
   w_0(L,K) &=& w(L,K), 
\eea
with 
\bba
    y(L,K) &=& -2(2L-2K-1)x(L,K),
             \label{y} \\
    w(L,K) &=&-\sq{\fr{8(2L-2K-1)(2L-2K)}{(2K-1)2K(2K+3)}}x(L,K).
             \label{w}
\eea

The condition that the $c^\dag e^\dag$ term vanishes determines 
the function $v_0$. 
Using the already derived functions, $y_0=y$ (\ref{y}) and $w_0=w$ (\ref{w}), 
and the crossing relation of type II (\ref{cross-II}) for $n=0$ 
with $J=L-1$, $J_1=L-K-1$ and $J_2=K$, in which the triangular conditions, 
\bb
    {}^0\oI^{L-1 N}_{L-K-1(M_1x_1);K M_2} =0 
           \label{0I} 
\ee
and so on, are taken into account, we find that $v_0(L,K)=v (L,K)$ with 
\bb
   v(L,K)= 2\sq{ \fr{(2K+1)(2L-2K-1)}{(2K-1)(2K+3)(2L-2K-3)(2L-2K+1)} }
             x \left( L, K+\half \right). 
           \label{v}
\ee

The $e^\dag e^\dag$ term now consists of only the $\D \cdot {}^0\oD$ part.   
Here, we consider the crossing relation of type III (\ref{cross-III}) for $n=0$ 
with $J=L-1$, $J_1=L-K-\fr{3}{2}$ and $J_2=K$.  
We then obtain the relation that consists of only the $\D$ and ${}^0\oD$ coefficients, 
because 
\bba 
   &&{}^0\oD^{L-1 N}_{L-K-2(M_1y_1),K(M_2y_2)}
      ={}^0\oD^{L-1 N}_{K-\half(M_1y_1),L-K-\fr{3}{2}(M_2y_2)}=0, 
                \\
   &&{}^0\oH^{L-1 N}_{L-K-\fr{3}{2}(M_1x_1);K(M_2y_2)}
      ={}^0\oH^{L-1 N}_{K(M_1x_1);L-K-\fr{3}{2}(M_2y_2)}=0,
                 \\
   &&{}^0\oG^{L-1 N}_{L-K-\fr{3}{2}(M_1y_1);K M_2}
      ={}^0\oG^{L-1 N}_{K(M_1y_1);L-K-\fr{3}{2} M_2}=0.
\eea 
Using this crossing relation and the function $v_0=v$ (\ref{v}), 
we find that the $e^\dag e^\dag$ term vanishes.

The $d^\dag c^\dag$ term consists of only the $\E \cdot {}^0\oE$ part. 
Using the crossing relation of type I (\ref{cross-I}) for $n=0$ and $y_0=y$ (\ref{y}), 
we can show that this term vanishes.

Thus, we obtain another type of $Q_M$ invariant operator with scalar index, denoted by
$\tilde{{\cal A}}^{\dag}_{L-1 N}$, as
\bba
  && \tilde{{\cal A}}^{\dag}_{L-1 N}
   = \sum_{K=1}^{L-1} \sum_{M_1,x_1}\sum_{M_2,x_2} \bar{x}(L,K)
      \E^{L-1N}_{L-K (M_1x_1), K(M_2,x_2)}
      \tc^{\dag}_{L-K(M_1x_1)} \tc^{\dag}_{K(M_2x_2)}
               \nonumber \\
  &&\qquad
     + \sum_{K=1}^{L-2} \sum_{M_1,x_1}\sum_{M_2,x_2} \bar{y}(L,K)
      \E^{L-1N}_{L-K-1 (M_1x_1), K(M_2,x_2)}
      \td^{\dag}_{L-K-1 (M_1x_1)} \tc^{\dag}_{K(M_2x_2)}
               \nonumber  \\
  &&\qquad
     +\sum_{K=1}^{L-\fr{3}{2}} \sum_{M_1,x_1}\sum_{M_2,y_2} \bar{w}(L,K)
      \H^{L-1N}_{L-K-\half (M_1x_1); K(M_2,y_2)}
      \tc^{\dag}_{L-K-\half (M_1x_1)} \te^{\dag}_{K(M_2y_2)}
               \nonumber \\
  &&\qquad
     +\sum_{K=1}^{L-2} \sum_{M_1,y_1}\sum_{M_2,y_2} \bar{v}^\pp (L,K)
      \D^{L-1N}_{L-K-1 (M_1y_1), K(M_2,y_2)}
      \te^{\dag}_{L-K-1(M_1y_1)} \te^{\dag}_{K(M_2y_2)} ,
             \nonumber \\ 
\eea
with integer $L (\geq 3)$. Here, the definitions ${}^0\oE =\E$ and ${}^0\oH =\H$, 
and the relation between $\D$ and ${}^0\oD$ (\ref{0D(J,K)}) are used. 
The function $v^\pp$ is then given by
\bb
   v^\pp (L,K) = -\half (2K-1)(2L-2K-3) v (L,K).
\ee
The functions with the bar ($\by$, $\bw$, and $\bv^\pp$) are defined as in 
(\ref{by}), (\ref{bw}) and (\ref{bv}), respectively.

For  $J=L-2$, all components in the operator (\ref{O}) contribute. 
In this case we find that we cannot make some terms in the commutator (\ref{Q-O}) vanish. 
For example, the $e^\dag e^\dag$ term is now given by a combination of  
the $\H \cdot {}^0\oH$ and $\D \cdot{}^0\oD$ parts. 
The crossing relation required in this case is of type III (\ref{cross-III}) 
for $n=0$ with $J=L-2$, $J_1=L-K-\fr{3}{2}$ and $J_2=K$.   
However, this relation also includes the non-vanishing $\G \cdot {}^0\oG$ part,  
so that we cannot make this term vanish unless the functions $u_0$ and $v_0$ vanish. 
From the other terms in the commutator, these functions must be non-vanishing, 
and thus we can show that for $J \leq L-2$, there is no operator that 
commutes with the charge $Q_M$.     
Thus, the building blocks with scalar index are given by two types of 
the operators, $A^\dag$ and ${\cal A}^\dag$. 

\paragraph{Building blocks with vector index $(n=1)$:}
{}From the triangular conditions of the Clebsch-Gordan coefficients, 
the operator with vector index is non-vanishing for $J \leq L-\half$. 
Let us consider the case of $J=L-\half$. Because 
\bba
   && {}^1\oE^{L-\half (Ny)}_{L-K-1(M_1x_1),K(M_2x_2)}
    ={}^1\oE^{L-\half (Ny)}_{L-K-2(M_1x_1),K(M_2x_2)}=0, \\
   && {}^1\oH^{L-\half (Ny)}_{L-K-\fr{3}{2}(M_1x_1);K(M_2y_2)}=0, \\
   && {}^1\oD^{L-\half (Ny)}_{L-K-1(M_1y_1),K(M_2y_2)}=0,   
\eea
only the terms including the functions $x_1$ and $w_1$ survive. 
Therefore, the four terms ($d^\dag d^\dag$, $e^\dag e^\dag$, $d^\dag c^\dag$ 
and $d^\dag e^\dag$) in the commutator (\ref{Q-O}) vanish trivially. 
Using the crossing relations of type I (\ref{cross-I}) 
and type II (\ref{cross-II}) for $n=1$, 
we find that the $c^\dag c^\dag$ and $c^\dag e^\dag$ terms vanish when 
\bba
    && x_1(L,K)=x(L,K), \\ 
    && w_1(L,K)=w(L,K), 
\eea
where $x$ and $w$ are given by Eqs.(\ref{x}) and (\ref{w}). 
Thus, we obtain a building block with vector index, 
denoted by $\tilde{B}^{\dag}_{L-\half (Ny)}$, as
\bba
  && \tilde{B}^{\dag}_{L-\half (Ny)}
   = \sum_{K=1}^{L-1} \sum_{M_1,x_1}\sum_{M_2,x_2} \bar{x}(L,K)
      ~{}^1\oE^{L-\half (Ny)}_{L-K (M_1x_1), K(M_2,x_2)}
      \tc^{\dag}_{L-K(M_1x_1)} \tc^{\dag}_{K(M_2x_2)}
               \nonumber \\
  &&\qquad
     +\sum_{K=1}^{L-\fr{3}{2}} \sum_{M_1,x_1}\sum_{M_2,y_2} \bar{w}(L,K)
      ~{}^1\oH^{L-\half (Ny)}_{L-K-\half (M_1x_1); K(M_2,y_2)}
      \tc^{\dag}_{L-K-\half (M_1x_1)} \te^{\dag}_{K(M_2y_2)},
               \nonumber \\
\eea
with integer $L (\geq 3)$.

   For $J \leq L-\fr{3}{2}$, we find that the operator does not commute with $Q_M$, 
as discussed before.

\paragraph{Building blocks with rank 2 tensor index $(n=2)$:}
The operator with rank 2 tensor index  
has non-vanishing components for $J \leq L-1$. 
For $J=L-1$, because of 
\bba
  && {}^2\oE^{L-1 (Nx)}_{L-K-1 (M_1x_1), K(M_2,x_2)}
   ={}^2\oE^{L-1 (Nx)}_{L-K-2 (M_1x_1), K(M_2,x_2)}=0, \\
  && {}^2\oH^{L-1 (Nx)}_{L-K-\fr{3}{2}(M_1x_1); K(M_2,y_2)}=0,
\eea
the terms with the functions $\bar{x}_2$, $\bar{w}_2$ and $\bar{v}_2$ survive. 
Therefore, we consider the three terms $c^\dag c^\dag$, $e^\dag e^\dag$ and $c^\dag e^\dag$ 
in the commutator (\ref{Q-O}).   
As in the case of the operator ${\cal A}^\dag$, using the crossing relations of 
type I (\ref{cross-I}) and type III (\ref{cross-III}) for $n=2$, 
we can make the $c^\dag c^\dag$ and $e^\dag e^\dag$ terms vanish 
when the functions $x_2$, $w_2$ and $v_2$ are given by $x$ (\ref{x}), 
$w$ (\ref{w}) and $v$ (\ref{v}), respectively.
However, we cannot make the $c^\dag e^\dag$ term vanish  
because the coresponding type II relation (\ref{cross-II}) for $n=2$ has 
an extra term including the non-vanishing coefficient ${}^2\oI^{L-1}_{L-K-1;K}$  
contrary to (\ref{0I}).    
Thus, the operator with $J=L-1$ does not commute with $Q_M$.

   In the same way, we can show that there is no $Q_M$-invariant operator with $J \leq L-2$. 
Thus, the building block with rank 2 tensor index is the only lowest positive-metric 
creation mode,
\bb
         c^{\dag}_{1(Nx)}.
\ee

\paragraph{Building blocks with rank 3 tensor index $(n=3)$:}
The operator with  rank 3 tensor index  
has non-vanishing components for $J \leq L-\half$. 
For $J=L-\half$, because of 
\bba
   && {}^3\oE^{L-\half (Nz)}_{L-K-1(M_1x_1),K(M_2x_2)}
    ={}^3\oE^{L-\half (Nz)}_{L-K-2(M_1x_1),K(M_2x_2)}=0, \\
   && {}^3\oH^{L-\half (Nz)}_{L-K-\fr{3}{2}(M_1x_1);K(M_2y_2)}=0, \\
   && {}^3\oD^{L-\half (Nz)}_{L-K-1(M_1y_1),K(M_2y_2)}=0,   
\eea
the terms with the functions $\bar{x}_3$ and $\bar{w}_3$ survive. 
Therefore, we consider only the  $c^\dag c^\dag$ and $c^\dag e^\dag$ terms 
in the commutator (\ref{Q-O}). 
As in the case of the operator $B^\dag$, using the crossing relation of type I (\ref{cross-I}) 
for $n=3$, we find that the $c^\dag c^\dag$ term vanishes when
\bba
     x_3(L,K)&=&x(L,K), \\
     w_3(L,K)&=&w(L,K),
\eea
where $x$ and $w$ are given by Eqs.(\ref{x}) and (\ref{w}). 
Also, using this $w_3$, we find that the $c^\dag e^\dag$ term vanishes 
due to the crossing relation of type II (\ref{cross-II}) for $n=3$.

Thus, we obtain a building block with rank 3 tensor index, 
denoted by $\tilde{D}^{\dag}_{L-\half (Nz)}$, as
\bba
  && \tilde{D}^{\dag}_{L-\half (Nz)}
   = \sum_{K=1}^{L-1} \sum_{M_1,x_1}\sum_{M_2,x_2} \bar{x}(L,K)
      ~{}^3\oE^{L-\half (Nz)}_{L-K (M_1x_1), K(M_2,x_2)}
      \tc^{\dag}_{L-K(M_1x_1)} \tc^{\dag}_{K(M_2x_2)}
               \nonumber \\
  &&\qquad
     +\sum_{K=1}^{L-\fr{3}{2}} \sum_{M_1,x_1}\sum_{M_2,y_2} \bar{w}(L,K)
      ~{}^3\oH^{L-\half (Nz)}_{L-K-\half (M_1x_1); K(M_2,y_2)}
      \tc^{\dag}_{L-K-\half (M_1x_1)} \te^{\dag}_{K(M_2y_2)},
               \nonumber \\
\eea
with integer $L (\geq 3)$.

For $J < L-\half$, the operator does not commute with $Q_M$. 

\paragraph{Building blocks with rank 4 tensor index $(n=4)$:}
The operator with  rank 4 tensor index  
has non-vanishing components for $J \leq L$. 
For $J=L$, because of 
\bba
    && {}^4\oE^{L(Nw)}_{L-K-1(M_1x_1),K(M_2x_2)}={}^4\oE^{L(Nw)}_{L-K-2(M_1x_1),K(M_2x_2)}=0, 
                         \\
    && {}^4\oH^{L(Nw)}_{L-K-\half (M_1x_1);K(M_2y_2)}
         ={}^4\oH^{L(Nw)}_{L-K-\fr{3}{2}(M_1x_1);K(M_2y_2)}=0,
                         \\
    && {}^4\oD^{L(Nw)}_{L-K-1(M_1y_1),K(M_2y_2)}=0, 
\eea
the terms with the function $\bar{x}_4$ survive, and therefore  
we cosider only the $c^\dag c^\dag$ term in the commutator (\ref{Q-O}). 
We easily find that, using the crossing relation of type I (\ref{cross-I}) for $n=4$,  
this term vanishes when  
\bb
     x_4(L,K)=x(L,K). 
\ee   
Thus, we obtain a building block with rank 4 tensor index, 
denoted by $\tilde{E}^{\dag}_{L (Nw)}$, as 
\bb
   \tilde{E}^{\dag}_{L (Nw)}
   = \sum_{K=1}^{L-1} \sum_{M_1,x_1}\sum_{M_2,x_2} \bar{x}(L,K)
      ~{}^4\oE^{L(Nw)}_{~L-K (M_1x_1), K(M_2,x_2)}
      \tc^{\dag}_{L-K(M_1x_1)} \tc^{\dag}_{K(M_2x_2)},
\ee
with integer $L (\geq 3)$.

    For $J=L-1$, because of 
\bba
   && {}^4\oE^{L-1 (Nw)}_{L-K-2(M_1x_1),K(M_2x_2)}=0, \\ 
   && {}^4\oH^{L-1 (Nw)}_{L-K-\fr{3}{2}(M_1x_1);K(M_2y_2)}=0, \\
   && {}^4\oD^{L-1 (Nw)}_{L-K-1(M_1y_1),K(M_2y_2)}=0,
\eea
the terms with the functions $\bar{x}_4$, $\bar{y}_4$ and $\bar{w}_4$ survive.  
Therefore, we consider the three terms $c^\dag c^\dag$, $d^\dag c^\dag$ and $c^\dag e^\dag$  
in the commutator (\ref{Q-O}).  
As in the case of the operator ${\cal A}^\dag$,  using the crossing relations of 
type I (\ref{cross-I}) and type II (\ref{cross-II}) for $n=4$ in company with 
the triangular condition, 
\bb
      {}^4\oI^{L-1(Nw)}_{L-K-1(M_1x_1);K M_2}=0,
\ee 
we find that the commutator vanishes when 
\bba
     x_4(L,K)&=&x(L,K), \\
     y_4(L,K)&=&y(L,K), \\ 
     w_4(L,K)&=&w(L,K). 
\eea
Thus, we obtain another type of building block with rank 4 tensor index, 
denoted by $\tilde{{\cal E}}^{\dag}_{L-1 (Nw)}$, as
\bba
  && \tilde{{\cal E}}^{\dag}_{L-1 (Nw)}
   = \sum_{K=1}^{L-1} \sum_{M_1,x_1}\sum_{M_2,x_2} \bar{x}(L,K)
      ~{}^4\oE^{L-1(Nw)}_{L-K (M_1x_1), K(M_2,x_2)}
      \tc^{\dag}_{L-K(M_1x_1)} \tc^{\dag}_{K(M_2x_2)}
               \nonumber \\
  &&\qquad
     + \sum_{K=1}^{L-2} \sum_{M_1,x_1}\sum_{M_2,x_2} \bar{y}(L,K)
      ~{}^4\oE^{L-1(Nw)}_{~L-K-1 (M_1x_1), K(M_2,x_2)}
      \td^{\dag}_{L-K-1 (M_1x_1)} \tc^{\dag}_{K(M_2x_2)}
               \nonumber  \\
  &&\qquad
     +\sum_{K=1}^{L-\fr{3}{2}} \sum_{M_1,x_1}\sum_{M_2,y_2} \bar{w}(L,K)
      ~{}^4\oH^{L-1(Nw)}_{~L-K-\half (M_1x_1); K(M_2,y_2)}
      \tc^{\dag}_{L-K-\half (M_1x_1)} \te^{\dag}_{K(M_2y_2)},
               \nonumber \\
\eea
with integer $L (\geq 3)$.

For $J \leq L-3$, there is no $Q_M$-invariant operator with rank 4 tensor index.

The building blocks in the traceless mode sector are summarized in Table 3. 
The operators without the tilde are defined 
by $O_{L(N\veps_n)}=\eps_N \tilde{O}_{L (-N\veps_n)}$.
Any $Q_M$-invariant state will be constructed from these building blocks. 
\begin{center}
\begin{tabular}{|c|ccccc|}  \hline
rank of tensor index & $0$  & $1$ & $2$ &$3$ & $4$    \\ \hline
creation op.  & $A^\dag_{LN}$ & $B^\dag_{L-\half(Ny)}$ & $c^\dag_{1 (Nx)}$ 
              & $D^\dag_{L-\half (Nz)}$ & $E^\dag_{L(Nw)}$  \\
              & ${\cal A}^\dag_{L-1 N}$ &  &  &  & ${\cal E}^\dag_{L-1 (Nw)}$  \\ 
level $(L \in {\bf Z}_{\geq 3})$ &  $2L$  & $2L$  & $2$ & $2L$ & $2L$       \\ \hline
\end{tabular} \\
Table 3: Building blocks in the traceless mode sector. 
\end{center}

\section{Building Blocks for the Conformal Field}
\setcounter{equation}{0}
\noindent

There is an essential difference between the conformal field and the other fields,   
which is that the conformal field has zero modes.  
The commutators of $Q_M$ and the zero modes are given by
\bba
     \left[ Q_M, \hat{q} \right] &=& -a_{\half M}, 
            \\
     \left[ Q_M, \hat{p} \right] &=& 0.     
\eea
The commutators with the conformal modes $\ta^{\dag}_{JM}$ are calculated as
\bb
    \left[ Q_M, \ta^{\dag}_{\half M_1} \right]
    = \left( \sq{2b_1}-i\hat{p} \right) \eps_{M_1} \dl_{M,-M_1}
\ee
and
\bb
   \left[ Q_M, \ta^{\dag}_{JM_1} \right]
   = \a \left( J-\half \right) \sum_{M_2} \eps_{M_1}
     \C^{\half M}_{J -M_1, J-\half M_2}\ta^{\dag}_{J-\half M_2}
\ee
for $J \geq 1$.  Also, the commutators with $\tb^{\dag}_{JM}$ are 
given by
\bba
   \left[ Q_M, \tb^{\dag}_{JM_1} \right]
   &=& -\gm (J) \sum_{M_2} \eps_{M_1}
     \C^{\half M}_{J -M_1, J+\half M_2}\ta^{\dag}_{J+\half M_2}
           \nonumber \\
   &&  -\b \left( J-\half \right) \sum_{M_2} \eps_{M_1}
     \C^{\half M}_{J -M_1, J-\half M_2}\tb^{\dag}_{J-\half M_2}
\eea
for $J\geq 0$.

The building blocks are constructed as done in the case of the tracceless mode.
The differences are that we here use the Clebsch-Gordan 
coefficient of type $\C$, and we take care on the zero-mode.
Then, we find two types of the building blocks with level $H=2L$,   
\bba
     \tilde{S}^{\dag}_{L N}
     &=& \sum_{K=\half}^{L-\half} \sum_{M_1,M_2} \bar{x}(L,K)
         \C^{L N}_{L-K M_1, K M_2}
         \ta^{\dag}_{L-K M_1} \ta^{\dag}_{K M_2}
               \nonumber \\
      && + \chi(\hat{p}) \ta^{\dag}_{L N}
\eea
and
\bba
     \tilde{{\cal S}}^{\dag}_{L-1 N}
     &=& \sum_{K=\half}^{L-\half} \sum_{M_1,M_2} \bar{x}(L,K)
        \C^{L-1 N}_{L-K M_1, K M_2}
        \ta^{\dag}_{L-K M_1} \ta^{\dag}_{K M_2}
            \nonumber  \\
      && +\sum_{K=\half}^{L-1} \sum_{M_1,M_2} \bar{y}(L,K)
        \C^{L-1 N}_{L-K-1 M_1, K M_2}
        \tb^{\dag}_{L-K-1 M_1} \ta^{\dag}_{K M_2}
             \nonumber \\
      && +\psi(\hat{p}) \tb^{\dag}_{L-1 N},
\eea 
with integer $L (\geq 1)$. 
Here, $\bar{x}$ and $\bar{y}$ are the same to the functions defined 
by Eqs.(\ref{x}) and (\ref{y}) in the traceless-mode sector.
The zero-mode dependent operators, $\chi$ and $\psi$, are given by
\bba
    \chi (\hat{p}) &=& \fr{1}{\sq{2(2L-1)(2L+1)}} \left( \sq{2b_1}-i\hat{p}\right),
                      \\
    \psi (\hat{p}) &=& -\sq{2} \left( \sq{2b_1}-i\hat{p} \right).
\eea

The building blocks in the conformal mode sector are summarized in Table 4.  
The operators without the tilde are defined by
$S_{L N} = \eps_N \tilde{S}_{L -N}$ and ${\cal S}_{L N} =\eps_N \tilde{{\cal S}}_{L -N}$.

\begin{center}
\begin{tabular}{|c|c|}  \hline
rank of tensor index & $0$  \\ \hline
creation op.          & $S^\dag_{LN}$  \\
                      & ${\cal S}^\dag_{L-1N}  $  \\ 
level $(L \in {\bf Z}_{\geq 1})$ &  $2L$    \\ \hline
\end{tabular} \\
Table 4: Building blocks in the conformal mode sector. 
\end{center}

\section{Physical States in a Non-critical 3-brane}
\setcounter{equation}{0}
\noindent

The physical state annihilated by all conformal charges is a conformally invariant vacuum, 
which is uniquely given by
\bb
   |\Om \rangle = \e^{-\sq{2b_1}\hat{q}} |0 \rangle 
                = \e^{-2b_1 \phi_0} |0 \rangle,
\ee 
where $\phi_0$ indicates the zero mode of the conformal field (\ref{conformal-mode}), and   
$|0 \rangle$ is the standard Fock vacuum with zero eigenvalue of $\hat{p}$ that vanishes 
when annihilation modes act.
The physical states are spanned by the Fock space generated on the conformally invariant 
vacuum. They must satisfy the conformal invariance conditions,
\bb
      Q_M |{\rm phys} \rangle =0
            \label{Q-condition}
\ee
and
\bb
      (H-4)|{\rm phys} \rangle = R_{MN} |{\rm phys} \rangle =0, 
           \label{H-R-conditions}
\ee 
where $-4$ comes from the ghost sector discussed in Appendix F, which indicates 
the number of the dimensions of the world-volume. 
As in the Gupta-Bleuler procedure, we do not impose a condition concerning $Q_M^\dag$.

The physical state is now decomposed into four sectors: scalar fields, 
vector fields, the traceless mode and the conformal mode. 
Each sector consists of the Hamiltonian eigenstates satisfying 
the condition (\ref{Q-condition}). Such states are constructed from 
the building blocks derived in the previous sections. 
Conditions (\ref{H-R-conditions}) are imposed last after combining all sectors.

First, consider the states that depends only on the zero mode of the conformal field. 
Such a state satisfying the $Q_M$ invariance condition (\ref{Q-condition}) is 
given by $|p,\Om\rangle = \e^{ip\hat{q}}|\Om \rangle = \e^{ip\sq{2b_1}\phi_0}|\Om \rangle$.  
This is the eigenstate of $\hat{p}$ with eigenvalue $p+i\sq{2b_1}$. 
The Hamiltonian condition in (\ref{H-R-conditions}) 
gives the equation $\half \left( p+i\sq{2b_1} \right)^2+b_1=4$, so that $p$ has a purely 
imaginary value, $-i\fr{\a_0}{\sq{2b_1}}$, with $\a_0=2b_1 \left( 1-\sq{1-\fr{4}{b_1}} \right)$.
Here, the fact that $b_1 >4$ is used, and the solution that $\a$ approaches 
the canonical value, $4$, in the classical limit, $b_1 \rightarrow \infty$, is selected. 
This state is, expressed by
\bb
         \e^{\a_0 \phi_0} |\Om \rangle, 
\ee
identified with the cosmological constant.

The general state satisfying the conditions of $Q_M$ (\ref{Q-condition}) 
and $R_{MN}$ in (\ref{H-R-conditions}) is constructed by acting with the building 
blocks on the state $|p,\Om \rangle$, in which all tensor indices are 
contracted out using the $SU(2)\times SU(2)$ Clebsch Gordan coefficients.    
This state is denoted by  $|n,p \rangle= F_n(\Phi^\dag, \cdots)|p,\Om \rangle$, 
where $n$ is the level of $F_n$.  The Hamiltonian condition gives the 
equation $\half \left( p+i\sq{2b_1} \right)^2+b_1 +n =4$. Solving this equation, 
we obtain the physical state
\bb
    F_n(\Phi^\dag, \cdots)\e^{\a_n \phi_0}|\Om \rangle 
\ee
with the conformal charge
\bb
   \a_n=2b_1 \left( 1-\sq{1-\fr{4-n}{b_1}} \right). 
\ee

Now, we construct the lower level states up to the level $6$. 
For $n=2$, there are two physical states,
\bb
     \Phi_{00}^\dag \e^{\a_2 \phi_0}|\Om \rangle 
\ee
and purely gravitational state
\bb
     {\cal S}^\dag_{00} \e^{\a_2 \phi_0}|\Om \rangle.    
\ee
The former corresponds to the diffeomorphism invariant field, $\int d^4 x \sq{-g} X^2$,  
and the latter is the scalar curvature, $\int d^4x \sq{-g}R$.

For $n=4$, the physical sates coupled to the matter fields are given by
\bb
    \left( \Phi_{00}^\dag \right)^2 |\Om \rangle, 
              \quad   
    \Phi_{00}^\dag {\cal S}^\dag_{00}  |\Om \rangle, 
              \quad
    \tq^\dag_{\half (Ny)} q^\dag_{\half (Ny)} |\Om \rangle, 
\ee
where $\a_4=0$ is taken into account. 
Here and below, the sums of the tensor indicies are omitted. 
These states correspond to the diffeomorphism invariant fields $\int d^4 x \sq{-g}X^4$, 
$\int d^4 x \sq{-g} RX^2$ and the square of the field strength of the vector 
field, $\int d^4x \sq{-g}F_{\mu\nu}F^{\mu\nu}$, respectively.
Furthermore, there are purely gravitational states,
\bb
      \tc_{1(Nx)}^\dag c_{1(Nx)}^\dag  |\Om \rangle, 
                 \quad     
      \left( {\cal S}^\dag_{00} \right)^2 |\Om \rangle, 
                 \quad
      \tilde{S}^\dag_{1N}S^\dag_{1N}  |\Om \rangle.
\ee
The first state corresponds to the square of the Weyl tensor, 
$\int d^4 x \sq{-g}C^2_{\mu\nu\lam\s}$, and the second is 
the square of the scalar curvature, $\int d^4 x \sq{-g} R^2$. 
The third is an independent diffeomorphism invariant field  
other than the first two fields.

For the level $n=6$, we obtain
\bba
   && \left( \Phi^\dag_{00} \right)^3 \e^{\a_6 \phi_0}|\Om \rangle, 
              \quad
      \left( \Phi^\dag_{00} \right)^2 {\cal S}^\dag_{00} \e^{\a_6 \phi_0}|\Om \rangle,
          \nonumber \\      
   && \Phi^\dag_{00} \left( {\cal S}^\dag_{00} \right)^2 \e^{\a_6 \phi_0}|\Om \rangle,
              \quad
      \Phi^\dag_{00}  \tilde{S}^\dag_{1N} S^\dag_{1N} \e^{\a_6 \phi_0}|\Om \rangle,
           \nonumber \\
   && \tilde{\Phi}^\dag_{1N} S^\dag_{1N} \e^{\a_6 \phi_0} |\Om \rangle, 
                   \quad
       \Phi_{00}^\dag \tq^\dag_{\half (Ny)} 
           q^\dag_{\half (Ny)}  \e^{\a_6 \phi_0}|\Om \rangle, 
             \nonumber \\
   &&  \Phi^\dag_{00} \tc^\dag_{\half (Nx)} c^\dag_{\half (Nx)} 
          \e^{\a_6 \phi_0}|\Om \rangle,
                 \quad
       \tq^\dag_{\half (Ny)} q^\dag_{\half (Ny)} 
         {\cal S}^\dag_{00} \e^{\a_6 \phi_0}|\Om \rangle,
             \nonumber \\
   && \D^{1N}_{\half (N_1y_1), \half (N_2y_2)} \tq^\dag_{\half (N_1y_1)} 
         \tq^\dag_{\half (N_2y_2)} S^\dag_{1N} \e^{\a_6 \phi_0} |\Om \rangle, 
             \nonumber \\
   && \F^{1(Nx)}_{\half (N_1y_1), \half (N_2y_2)} \tq^\dag_{\half (N_1y_1)} 
         \tq^\dag_{\half (N_2y_2)} c^\dag_{1(Nx)} \e^{\a_6 \phi_0} |\Om \rangle, 
            \nonumber \\
   && \G^{1N_2}_{\half (Ny);1N_1} \tq^\dag_{\half (Ny)} 
         \tilde{S}^\dag_{1N_1} S^\dag_{1N_2} \e^{\a_6 \phi_0} |\Om \rangle, 
             \nonumber \\   
   && \H^{1N_3}_{1 (N_1x_1); \half (N_2y_2)} \tq^\dag_{\half (N_2y_2)} 
          \tc^\dag_{1(N_1x_1)} S^\dag_{1N_3} \e^{\a_6 \phi_0} |\Om \rangle,
             \nonumber \\
   && {}^1\oE^{\half (Ny)}_{1 (N_1x_1), 1 (N_2x_2)} q^\dag_{\half (Ny)} 
         \tc^\dag_{1 (N_1x_1)} \tc^\dag_{1 (N_2x_2)}  \e^{\a_6 \phi_0} |\Om \rangle  
\eea
and purely gravitational states
\bba
    && \left( {\cal S}^\dag_{00} \right)^3 \e^{\a_6 \phi_0}|\Om \rangle, 
               \quad
       {\cal S}^\dag_{00} \tilde{S}^\dag_{1N} S^\dag_{1N} \e^{\a_6 \phi_0}|\Om \rangle, 
               \nonumber  \\
    && \tilde{S}^\dag_{1N} {\cal S}^\dag_{1N} \e^{\a_6 \phi_0}|\Om \rangle,
               \quad 
       \tc^\dag_{1(Nx)} c^\dag_{1(Nx)} {\cal S}^\dag_{00} \e^{\a_6 \phi_0}|\Om \rangle,  
         \nonumber \\
    && \C^{1N_3}_{1 N_1,1N_2} \tilde{S}^\dag_{1N_1} \tilde{S}^\dag_{1N_2} 
          S^\dag_{1N_3} \e^{\a_6 \phi_0}|\Om \rangle, 
            \nonumber \\
    && \I^{1N_2}_{1(Nx);1N_1} \tc^\dag_{1(Nx)} \tilde{S}^\dag_{1N_1} 
          S^\dag_{1N_2} \e^{\a_6 \phi_0}|\Om \rangle.  
               \nonumber \\
    && \E^{1N}_{1(N_1x_1),1(N_2x_2)} \tc^\dag_{1(N_1x_1)}\tc^\dag_{1(N_2x_2)} 
          S^\dag_{1N} \e^{\a_6 \phi_0}|\Om \rangle, 
               \nonumber \\
    && {}^2\oE^{1(N_3x_3)}_{1(N_1x_1),1(N_2x_2)}
       \tc^\dag_{1(N_1x_1)}\tc^\dag_{1(N_2x_2)} c^\dag_{1(N_3x_3)}
         \e^{\a_6 \phi_0}|\Om \rangle, 
\eea
In the same way we can also construct higher level states, but the classification 
of them becomes complicated.

\section{Conclusions and Discussion}
\setcounter{equation}{0}
\noindent

In this paper we systematically constructed and classified the physical states 
in a world-volume model of a non-critical 3-brane on $R\times S^3$ 
at very high energies beyond the Planck mass scale. 
At this energy, the conformal invariance, reflecting the background-metric independence, 
becomes exact, and thus the dynamics is described by CFT$_4$. 
Therefore, the physical states must satisfy the conformal invariance conditions, 
(\ref{Q-condition}) and (\ref{H-R-conditions}).

The physical states are decomposed into four sectors: scalar fields, vector fields, 
the traceless mode and the conformal mode. We discussed four sectors separately, 
and then combined them last.  
Each sector consists of the Hamiltonian eigenstates invariant under the special 
conformal transformations, $Q_M$, which belong to certain representations 
of the rotation group on $S^3$.     
These eigenstates further factorize into the $Q_M$-invariant building blocks 
classified in finite types. 
The physical state was constructed by combining such eigenstates, 
and contracting out all of their tensor indices appropriately 
in a rotation invariant way using the $SU(2)\times SU(2)$ Clebsch-Gordan coefficients. 
The Hamiltonian condition was imposed by adjusting the zero-mode momentum 
of the conformal field with purely imaginary eigenvalue.

There is an essential difference between the conformal mode sector and the other three sectors: scalar fields, vector fields and the traceless mode. 
The conformal mode sector is not normalizable because of the purely imaginary 
eigenvalue of the zero-mode momentum, while the other three sectors are normalizable.  
This world-volume model seems to be in the same universarity class as the four-dimensional 
simplicial quantum gravity, namely the dynamical triangulation approach 
to four-dimensional random surfaces~\cite{hey}. 
The partition function of this lattice model is given 
by a grand canonical ensemble in the number of 4-simplices. 
This fact will be related to the non-normalizability of the conformal mode sector.
The other normalizable sectors are regarded as canonical ensembles on random surfaces.

As an impact to spacetime physics, this world-volume model 
gives a dynamical scenario of inflation consistent with 
observations of the cosmic microwave background anisotropies~\cite{hy}.

\newpage

\begin{center}
{\Large {\bf Appendix}}
\end{center}

\appendix
\section{The Metric on $R\times S^3$}
\setcounter{equation}{0}
\noindent

The metric on $R \times S^3$ is parametrized as 
\bba
     d{\hat s}^2_{R\times S^3} &=& \hg_{\mu\nu}dx^{\mu}dx^{\nu}
                  =-dt^2 + \hgm_{ij}dx^i dx^j
            \nonumber  \\
                 &=&-dt^2 + \fr{1}{4} (d\a^2 +d\b^2 +d\gm^2 +2 \cos \b d\a d\gm ),
           \label{metric}
\eea
where $t$ is the time  and $x^i=(\a,\b,\gm)$, with $i=1,2,3$,
are the Euler's angles.
Then, $\hR_{0\mu\nu\lam}=\hR_{0\mu}=0$,
$\hR_{ijkl}=(\hgm_{ik}\hgm_{jl}-\hgm_{il}\hgm_{jk})$, $\hR_{ij}=2\hgm_{ij}$ and $\hR=6$.

The volume element on unit $S^3$ is 
\bb
      d\Om_3 =d^3 x \sq{\hgm} =\fr{1}{8}\sin \b d\a d\b d\gm,
\ee
and the volume is ${\rm Vol}(S^3) =2\pi^2$.

\section{$SU(2) \times SU(2)$ Clebsch-Gordan Coefficients of Types, 
$\C$, $\D$, $\E$, $\G$ and $\H$}
\setcounter{equation}{0}
\noindent

The $SU(2) \times SU(2)$ Clebsch-Gordan coefficients are defined by the integrals 
of three products of ST$^2$ tensor harmonics.
Here, we give the basic coefficients of types, ($\C$, $\D$, $\E$, $\G$ and $\H$), 
calculated in Ref.\cite{hh}.

\paragraph{Type $\C$}
\bba
     \C^{JM}_{J_1M_1,J_2M_2}
      &=& \sq{{\rm Vol}(S^3)}
         \int_{S^3} d\Om_3 Y^*_{JM}Y_{J_1M_1}Y_{J_2M_2}
             \nonumber \\
      &=& \sq{\fr{(2J_1+1)(2J_2+1)}{2J+1}} C^{Jm}_{J_1m_1,J_2m_2}
                           C^{J\prm}_{J_1\prm_1,J_2\prm_2}, 
\eea 
where $M=M_1 +M_2$ and
\bb
   |J_1 -J_2| \leq J \leq J_1 +J_2,
      \label{triangular-C}
\ee
with integer $J+J_1+J_2$.

\paragraph{Type $\D$ $(y=\pm\half)$}
\bba
  && \D^{JM}_{J_1(M_1y_1),J_2(M_2y_2)}
     = \sq{{\rm Vol}(S^3)}
         \int_{S^3} d\Om_3 Y^*_{JM}Y^i_{J_1(M_1y_1)}Y_{i J_2(M_2y_2)}
              \nonumber \\
  && = -\sq{ \fr{2J_1(2J_1+1)(2J_1+2)2J_2(2J_2+1)(2J_2+2)}{2J+1} }
              \nonumber \\
  && \quad \times
     \left\{ \begin{array}{ccc}
              J   & J_1     & J_2 \\
            \half & J_2+y_2 & J_1+y_1
            \end{array} \right\}
     \left\{ \begin{array}{ccc}
              J   & J_1     & J_2 \\
            \half & J_2-y_2 & J_1-y_1
            \end{array} \right\}
                 \nonumber \\
   && \quad \times
      C^{Jm}_{J_1+y_1 m_1,J_2+y_2 m_2}
      C^{J\prm}_{J_1-y_1 \prm_1,J_2-y_2 \prm_2}, 
\eea
where $M=M_1 +M_2$ and
\bb
     |J_1 -J_2| \leq J \leq J_1 +J_2 ,
         \label{triangular-D}
\ee
with integer $J+J_1+J_2$. 
The lower (upper) equality is satulated at $y_1 =y_2$ $(y_1 \neq y_2)$.

\paragraph{Type $\E$ $(x=\pm 1)$}
\bba
   && \E^{JM}_{J_1(M_1x_1),J_2(M_2x_2)}
     =  \sq{{\rm Vol}(S^3)}
         \int_{S^3} d\Om_3 Y^*_{JM}Y^{ij}_{J_1(M_1x_1)}Y_{ij J_2(M_2x_2)}
            \nonumber  \\
  && = \sq{ \fr{(2J_1-1)(2J_1+1)(2J_1+3)(2J_2-1)(2J_2+1)(2J_2+3)}{2J+1} }
            \nonumber \\
  && \quad \times
     \left\{ \begin{array}{ccc}
              J   & J_1     & J_2 \\
              1 & J_2+x_2 & J_1+x_1
            \end{array} \right\}
     \left\{ \begin{array}{ccc}
              J   & J_1     & J_2 \\
              1 & J_2-x_2 & J_1-x_1
            \end{array} \right\}
                  \nonumber \\
  && \quad \times
      C^{Jm}_{J_1+x_1 m_1,J_2+x_2 m_2}
      C^{J\prm}_{J_1-x_1 \prm_1,J_2-x_2 \prm_2},
\eea
where $M=M_1+M_2$ and 
\bb
     |J_1 -J_2| \leq J \leq J_1 +J_2 ,
        \label{triangular-E}
\ee
with integer $J+J_1+J_2$. 
The lower (upper) equality is satulated at $x_1 =x_2$ $(x_1 \neq x_2)$.

\paragraph{Type $\G$ $(y=\pm \half)$}
\bba
  && \G^{JM}_{J_1(M_1y_1);J_2 M_2}
     = \sq{{\rm Vol}(S^3)}
         \int_{S^3} d\Om_3 Y^*_{JM}Y^i_{J_1(M_1y_1)}\hnabla_i Y_{J_2 M_2}
           \nonumber \\
  && = -\fr{1}{2\sq{2}} \sq{ \fr{2J_1(2J_1+1)(2J_1+2)(2J_2+1)}{2J+1} }
           \sum_{K=J_2 \pm \half} 2K(2K+1)(2K+2)
              \nonumber   \\
  && \quad \times
      \left\{ \begin{array}{ccc}
              J   &   J_1   & K \\
            \half &   J_2   & J_1+\half
            \end{array} \right\}
     \left\{ \begin{array}{ccc}
              J   &  J_1    & K \\
            \half &  J_2    & J_1-\half
            \end{array} \right\}
      C^{Jm}_{J_1+y_1 m_1,J_2 m_2}
      C^{J\prm}_{J_1-y_1 \prm_1,J_2 \prm_2},
           \nonumber \\ 
  &&
\eea
where $M=M_1+M_2$ and
\bb
    |J_1 -J_2|+\half \leq J \leq J_1 +J_2 -\half,
         \label{triangular-G}
\ee
with half integer $J+J_1+J_2$. 

\paragraph{Type $\H$ $(x=\pm 1, y=\pm\half)$}
\bba
   && \H^{JM}_{J_1(M_1x_1);J_2(M_2y_2)}
      = \sq{{\rm Vol}(S^3)}
         \int_{S^3} d\Om_3 Y^*_{JM}Y^{ij}_{J_1(M_1x_1)}\hnabla_i Y_{j J_2(M_2y_2)}
             \nonumber \\
  && = -\fr{3}{2\sq{2}}\sq{ \fr{(2J_1-1)(2J_1+1)(2J_1+3)2J_2(2J_2+1)(2J_2+2)}{2J+1} }
              \nonumber \\
  && \quad \times
       \sum_{K=J_2 \pm \half} 2K(2K+1)(2K+2)
            \nonumber  \\
  && \quad \times
      \left\{ \begin{array}{ccc}
              K   &   1    & J_2+y_2 \\
            \half &  J_2   & \half
            \end{array} \right\}
     \left\{ \begin{array}{ccc}
              K   &  1   & J_2-y_2 \\
            \half & J_2  & \half
            \end{array} \right\}
     \left\{ \begin{array}{ccc}
              J  & J_1+x_1  & J_2+y_2 \\
              1  &    K     & J_1
            \end{array} \right\}
                \nonumber \\
   && \quad \times
     \left\{ \begin{array}{ccc}
              J & J_1-x_1  & J_2-y_2 \\
              1 &    K     & J_1
            \end{array} \right\}
      C^{Jm}_{J_1+x_1 m_1,J_2+y_2 m_2}
      C^{J\prm}_{J_1-x_1 \prm_1,J_2-y_2 \prm_2},
\eea
where $M=M_1+M_2$ and 
\bb
    |J_1 -J_2|+\half \leq J \leq J_1 +J_2 -\half,
      \label{triangular-H}
\ee
with half integer $J+J_1+J_2$. 
The lower (upper) equality is satulated at $x_1 =2y_2$ $(x_1 \neq 2y_2)$.

\section{Generalized Forms of $SU(2) \times SU(2)$ Clebsch-Gordan Coefficients}
\setcounter{equation}{0}
\noindent

General forms of $SU(2)\times SU(2)$ Clebsch-Gordan coefficients are 
defined by the integrals of scalar quantities constructed from three products 
of ST$^2$ tensor harmonics, 
\bba
   CG^{J(M\eps_n)}_{J_1(M_1\veps_{n_1}); J_2(M_2\veps_{n_2})}
   &\sim& \int_{S^3} d\Om_3 {\bf Sc}\left\{
           Y^{i_1 \cdots i_n *}_{J(M\veps_n)} \cdot
           Y^{j_1 \cdots j_{n_1}}_{J_1(M_1 \veps_{n_1})} \cdot
           Y^{k_1 \cdots k_{n_2}}_{J_2(M_2 \veps_{n_2})}
           \right\}
              \nonumber \\
    &\propto&  C^{J+\veps_n m}_{J_1+\veps_{n_1} m_1, J_2+\veps_{n_2} m_2}
       C^{J-\veps_n \prm}_{J_1-\veps_{n_1} \prm_1, J_2-\veps_{n_2} \prm_2},
               \label{CG-general}
\eea
where ${\bf Sc} \{ ~\}$ indicates the operation making the product  
a scalar quantity by contracting out tensor indices, $i$, $j$ and $k$, 
and, if necessary, inserting derivatives appropriately. 
This coefficient has a non-vanishing value when $M=M_1+M_2$ and  
the triangular conditions among $J$, $J_1$ and $J_2$ are satisfied, which are obtained  
from the conditions that two triangular conditions with respect 
to the left and right standard $SU(2)$ Clebsch-Gordan coefficients 
must be satisfied simultaneously.

Here, we define the generalized $SU(2)\times SU(2)$ Clebsch-Gordan 
coefficients, (${}^n\oE$, ${}^n\oH$, ${}^n\oD$, ${}^n\oG$ and ${}^n\oI$), 
used in the text to classify $Q_M$-invariant operators. 
The generalized  coefficients with the rank 4 tensor index
in the $J$ component are defined by 
\bba
   && {}^4\oE^{J (Mw)}_{J_1(M_1x_1), J_2(M_2x_2)}
      = \sq{{\rm Vol}(S^3)} \int_{S^3} d\Om_3 Y^{ijkl*}_{J(Mw)}
            Y_{ij J_1(M_1x_1)} Y_{kl J_2(M_2x_2)},
                   \nonumber \\
   && {}^4\oH^{J (Mw)}_{J_1(M_1x_1); J_2(M_2y_2)}
      = \sq{{\rm Vol}(S^3)} \int_{S^3} d\Om_3 Y^{ijkl*}_{J(Mw)}
         Y_{ij J_1(M_1x_1)} \hnabla_{(k}Y_{l) J_2(M_2y_2)},
                   \nonumber \\
   && {}^4\oD^{J (Mw)}_{J_1(M_1y_1), J_2(M_2y_2)}
      = \sq{{\rm Vol}(S^3)} \int_{S^3} d\Om_3 Y^{ijkl*}_{J(Mw)}
         \hnabla_{(i}Y_{j) J_1(M_1y_1)} \hnabla_{(k}Y_{l) J_2(M_2y_2)},
                   \nonumber \\
   && {}^4\oI^{J (Mw)}_{J_1(M_1x_1); J_2 M_2}
      = \sq{{\rm Vol}(S^3)} \int_{S^3} d\Om_3 Y^{ijkl*}_{J(Mw)}
         Y_{ij J_1(M_1x_1)} \hnabla_{(k} \hnabla_{l)}Y_{J_2 M_2},
                   \nonumber \\
   && {}^4\oG^{J (Mw)}_{J_1(M_1y_1); J_2 M_2}
      = \sq{{\rm Vol}(S^3)} \int_{S^3} d\Om_3 Y^{ijkl*}_{J(Mw)}
         \hnabla_{(i}Y_{j) J_1(M_1y_1)} \hnabla_{(k} \hnabla_{l)}Y_{J_2 M_2}.
             \label{general-4}
\eea

The generalized coefficients with the rank 3 tensor index
in the $J$ component, ${}^3\oE^{J (Mz)}_{J_1(M_1x_1), J_2(M_2x_2)}$,
${}^3\oH^{J (Mz)}_{J_1(M_1x_1); J_2(M_2y_2)}$, ${}^3\oD^{J (Mz)}_{J_1(M_1y_1), J_2(M_2y_2)}$,
${}^3\oI^{J (Mz)}_{J_1(M_1x_1); J_2 M_2}$ and ${}^3\oG^{J (Mz)}_{J_1(M_1y_1); J_2 M_2}$, are
defined by replacing $Y^{ijkl}_{J (Mw)}$ in (\ref{general-4}) 
with $\hnabla^{(i} Y^{jkl)}_{J(Mz)}$.

The generalized coefficients with the rank 2 tensor index in the $J$ component 
are defined by
\bba
   && {}^2\oE^{J (Mx)}_{J_1(M_1x_1), J_2(M_2x_2)}
      = \sq{{\rm Vol}(S^3)} \int_{S^3} d\Om_3 Y^{j*}_{i J(Mx)}
            Y^{k i}_{J_1(M_1x_1)} Y_{kj J_2(M_2x_2)},
                  \nonumber  \\
   && {}^2\oH^{J (Mx)}_{J_1(M_1x_1); J_2(M_2y_2)}
      = \sq{{\rm Vol}(S^3)} \int_{S^3} d\Om_3 Y^{j*}_{i J(Mx)}
         Y^{k i}_{J_1(M_1x_1)} \hnabla_{(k}Y_{j) J_2(M_2y_2)},
                  \nonumber  \\
   && {}^2\oD^{J (Mx)}_{J_1(M_1y_1), J_2(M_2y_2)}
      = \sq{{\rm Vol}(S^3)} \int_{S^3} d\Om_3 Y^{j*}_{i J(Mx)}
         \hnabla^{(k}Y^{i)}_{J_1(M_1y_1)} \hnabla_{(k}Y_{j) J_2(M_2y_2)},
                   \nonumber \\
   && {}^2\oI^{J (Mx)}_{J_1(M_1x_1); J_2 M_2}
      = \sq{{\rm Vol}(S^3)} \int_{S^3} d\Om_3 Y^{j*}_{i J(Mx)}
         Y^{k i}_{J_1(M_1x_1)} 
               \nonumber \\
   && \qquad\qquad\qquad\qquad\qquad\qquad\qquad\quad \times 
           \left( \hnabla_{(k} \hnabla_{j)} 
                            -\fr{1}{3}\hgm_{kj}\hnabla^2 \right) Y_{J_2 M_2},
                   \nonumber \\
   && {}^2\oG^{J (Mx)}_{J_1(M_1y_1); J_2 M_2}
      = \sq{{\rm Vol}(S^3)} \int_{S^3} d\Om_3 Y^{j*}_{i J(Mx)}
         \hnabla^{(k}Y^{i)}_{J_1(M_1y_1)} 
                  \nonumber  \\
   && \qquad\qquad\qquad\qquad\qquad\qquad\qquad\quad~ \times 
          \left( \hnabla_{(k} \hnabla_{j)} 
                         -\fr{1}{3}\hgm_{kj}\hnabla^2 \right) Y_{J_2 M_2}.
           \label{general-2}
\eea

The generalized coefficients with the vector index in the $J$ component,
${}^1\oE^{J (My)}_{J_1(M_1x_1), J_2(M_2x_2)}$,
${}^1\oH^{J (My)}_{J_1(M_1x_1); J_2(M_2y_2)}$, ${}^1\oD^{J (My)}_{J_1(M_1y_1), J_2(M_2y_2)}$,
${}^1\oI^{J (My)}_{J_1(M_1x_1); J_2 M_2}$ and ${}^1\oG^{J (My)}_{J_1(M_1y_1); J_2 M_2}$, are
defined by replacing $Y^{ij}_{J (Mx)}$ in (\ref{general-2}) 
with $\hnabla^{(i} Y^{j)}_{J(My)}$.

The generalized coefficients with the scalar index 
in the $J$ component are defined as
\bba
   && {}^0\oE^{J M}_{J_1(M_1x_1), J_2(M_2x_2)}
      = \E^{J M}_{J_1(M_1x_1), J_2(M_2x_2)},
                  \nonumber  \\
   && {}^0\oH^{J M}_{J_1(M_1x_1); J_2(M_2y_2)}
      = \H^{J M}_{J_1(M_1x_1); J_2(M_2y_2)},
                  \nonumber  \\
   && {}^0\oD^{J M}_{J_1(M_1y_1), J_2(M_2y_2)}
      = \sq{{\rm Vol}(S^3)} \int_{S^3} d\Om_3 Y^*_{J M}
         \hnabla^{(i}Y^{j)}_{J_1(M_1y_1)} \hnabla_{(i}Y_{j) J_2(M_2y_2)},
                   \nonumber \\
   && {}^0\oI^{J M}_{J_1(M_1x_1); J_2 M_2}
      = \I^{J M}_{J_1(M_1x_1); J_2 M_2} 
                 \nonumber  \\
   && \qquad\qquad\quad~~~
      = \sq{{\rm Vol}(S^3)} \int_{S^3} d\Om_3 Y^*_{J M}
         Y^{ij}_{J_1(M_1x_1)} \hnabla_{(i} \hnabla_{j)}Y_{J_2 M_2},
                   \nonumber \\
   && {}^0\oG^{J M}_{J_1(M_1y_1); J_2 M_2}
      = \sq{{\rm Vol}(S^3)} \int_{S^3} d\Om_3 Y^*_{J M}
         \hnabla^{(i}Y^{j)}_{J_1(M_1y_1)} \hnabla_{(i} \hnabla_{j)}Y_{J_2 M_2}.
\eea

The triangular conditions of these generalized Clebsch-Gordan coefficients are  
obtained from expression (\ref{CG-general}). 
The non-vanishing condition used in the text are summarized as
\bba
   {}^n\oE^J_{J_1,J_2} &:& J \leq J_1+J_2-\fr{n}{2} \quad (n\leq 2), 
               \nonumber  \\
                   &&  J \leq J_1+J_2- \left| \fr{n}{2}-2 \right| \quad (n \geq 2), 
           \label{triangular-oE}   \\
   {}^n \oH^J_{J_1;J_2} &:& J \leq J_1+J_2-\left| \fr{n}{2}-\half \right| \quad (n\leq 2),
                \nonumber \\
                  && J \leq J_1+J_2-\left| \fr{n}{2}-\fr{3}{2} \right| \quad (n\geq 2),
           \label{triangular-oH}   \\
   {}^n\oD^J_{J_1,J_2} &:& J \leq J_1+J_2-\fr{n}{2} \quad (n\leq 1),
                \nonumber \\
                   &&J \leq J_1+J_2-\left| \fr{n}{2}-1 \right| \quad (n\geq 1),
           \label{triangular-oD}   \\
   {}^n\oI^J_{J_1;J_2} &:& J \leq J_1+J_2-\left| \fr{n}{2}-1 \right| ,
           \label{triangular-oI}   \\
   {}^n\oG^J_{J_1;J_2} &:&  J \leq J_1+J_2-\left| \fr{n}{2}-\half \right| .           
           \label{triangular-oG}
\eea
The generalization of these coefficients to the cases with the higher rank index $n >4$ 
is straightforward, and then these conditions will be effective for such cases.

The coefficient ${}^0\oD$ can be expressed as
\bba
    && {}^0\oD^{J M}_{J_1(M_1y_1), J_2(M_2y_2)}
          \nonumber  \\
    && = \left\{ J_1(J_1+1)+J_2(J_2+1)-J(J+1)-\fr{3}{2} \right\}
        \D^{JM}_{J_1(M_1y_1), J_2(M_2y_2)}
             \nonumber \\
    && \quad + \half \tilde{\D}^{JM}_{J_1(M_1y_1), J_2(M_2y_2)},
            \label{0D}
\eea
where
\bb
   \tilde{\D}^{JM}_{J_1(M_1y_1), J_2(M_2y_2)}
   = \sq{{\rm Vol}(S^3)} \int_{S^3} d\Om_3 \left( \hnabla_i \hnabla_j Y^*_{JM} \right)
     Y^i_{J_1 (M_1y_1)} Y^j_{J_2 (M_2y_2)}.
            \label{tD}
\ee

The coefficient ${}^0\oG$ can be simplified in the form
\bb
    {}^0\oG^{J M}_{J_1(M_1y_1); J_2 M_2}
    = 2\left\{ J_1(J_1+1)+J_2(J_2+1)-J(J+1)-\fr{3}{4} \right\}
      \G^{J M}_{J_1(M_1y_1); J_2 M_2}.
         \label{0G}
\ee

\section{Relations between $\D$ and ${}^0\oD$}
\setcounter{equation}{0}
\noindent

The relations between the $\D^J_{J_1,J_2}$ and ${}^0\oD^J_{J_1,J_2}$ coefficients 
used in the text are derived here. 
The ${}^0\oD$ coefficient is expressed using $\tilde{\D}$ in (\ref{0D}). 
In the case of $J=\half$, 
because of $\hnabla_i \hnabla_j Y_{\half M}=-\hat{\gm}_{ij}Y_{\half M}$, 
we obtain the relation
\bb
   \tilde{\D}^{\half M}_{J_1(M_1y_1), J_2(M_2y_2)}
   =-\D^{\half M}_{J_1(M_1y_1), J_2(M_2y_2)}.
          \label{tD1/2}
\ee  
{}From this we obtain 
\bb
    {}^0\oD^{\half M}_{J_1(M_1y_1), J_2(M_2y_2)}
    = \left\{ J_1(J_1+1)+J_2(J_2+1)-\fr{11}{4} \right\}
        \D^{\half M}_{J_1(M_1y_1), J_2(M_2y_2)}.
          \label{0D1/2}
\ee

To discuss  more general cases used in Sect.5, we consider two types of 
crossing relations constructed from the $\D$ and $\tilde{\D}$ coefficients.
The one is the relation used to obtain the building blocks for the vector field in Sect.4,
\bba
   && \sum_{V,y} \eps_V \D^{\half M}_{J-K-\half (M_1y_1),J-K(-Vy)} 
                    \D^{JN}_{J-K(Vy),K(M_2y_2)} 
               \nonumber \\
   && = \sum_{V,y} \eps_V \D^{\half M}_{K (M_2y_2),K+\half(-Vy)} 
                    \D^{JN}_{K+\half(Vy),J-K-\half(M_1y_1)}.
               \label{D-D}
\eea
This is the special case of the crossing relation (\ref{cross-DD}) 
with $J_1=J-K-\half$ and $J_2=K$, where $J \geq 1$.
The other is the crossing relation derived from the integral
\bb
     \int_{S^3} d\Om_3 Y^*_{\half M} \hnabla^{[i}Y^{j]}_{J_1(M_1y_1)} 
     \hnabla_{[i}Y_{j] J_2(M_2y_2)} Y^*_{J N} .
\ee 
Because of the anti-symmetric property, the product expansion has the form 
\bba
   && Y^*_{\half M} \hnabla^{[i}Y^{j]}_{J_1 (M_1y_1)} 
          \nonumber \\
   && = -\fr{1}{\sq{{\rm Vol}(S^3)}} \sum_{I=J_1 \pm \half} \sum_{V,y} \fr{2}{(2I+1)^2} 
         \left\{ I(I+1)+J_1(J_1+1)+\fr{1}{4} \right\} 
           \nonumber \\
  && \qquad\qquad\qquad\qquad\qquad \times 
              \eps_V \D^{\half M}_{J_1(M_1y_1),I(-Vy)}  
        \hnabla^{[i}Y^{j]}_{I (Vy)} ,
\eea  
where Eq.(\ref{tD1/2}) is used. From this, 
we can obtain the crossing relation only with the $\D$ and $\tilde{\D}$ coefficients.  
For the case of $J_1=J-K-\half$ and $J_2=K$, it has the form
\bba
  && \fr{2J-2K}{2J-2K+1}  \sum_{V,y} 
          \eps_V \D^{\half M}_{J-K-\half (M_1 y_1), J-K(-V y)} 
         \nonumber \\ 
  && \qquad \times
       \left[ \left\{ -2K(2J-2K)+1 \right\} \D^{J N}_{J-K (V y), K (M_2 y_2)} 
           -\tilde{\D}^{J N}_{J-K (V y), K (M_2 y_2)} \right] 
         \nonumber  \\
  && = \fr{2K+1}{2K+2} \sum_{V,y}  
          \eps_V \D^{\half M}_{K (M_2 y_2), K+\half (-V y)} 
         \nonumber \\ 
  && \qquad \times
       \Bigl[ \left\{ -(2K+1)(2J-2K-1)+1 \right\} \D^{J N}_{K+\half (V y), J-K-\half (M_1 y_1)} 
         \nonumber \\
  && \qquad\qquad\qquad\qquad
           -\tilde{\D}^{J N}_{K+\half (V y), J-K-\half (M_1 y_1)} \Bigr]. 
         \label{D-tD}
\eea

These two crossing relations, (\ref{D-D}) and (\ref{D-tD}), should be equivalent. 
We here assume 
\bb
   \tilde{\D}^{JM}_{J-K(M_1y_1),K(M_2y_2)}=A(J,K)\D^{JN}_{J-K(M_1y_1),K(M_2y_2)}
          \label{tD(J,K)}
\ee 
with $A(J,K)=A(J,J-K)$ and $J \geq 1$. Substituting this relation into Eq.(\ref{D-tD}) 
and comparing with Eq.(\ref{D-D}), 
we obtain the recursion relation
\bb
     \fr{2J-2K}{2J-2K+1}B(J,K)=\fr{2K+1}{2K+2}B \left( J,K+\half \right),
\ee
where $B(J,K)=-2K(2J-2K)+1-A(J,K)$. 
Using the initial condition $A(J,\half)=2J+2$ easily calculated from the definition, 
we can solve the recusion relation, and thus we obtain the $K$ independent value, 
\bb
     A(J,K)=2J+2. 
       \label{A(J,K)}
\ee
{}From equation (\ref{0D}) and this result, we obtain the relation
\bb
    {}^0\oD^{JM}_{J-K(M_1y_1),K(M_2y_2)} 
    = -\half (2K-1)(2J-2K-1) \D^{JM}_{J-K(M_1y_1),K(M_2y_2)}. 
             \label{0D(J,K)}
\ee

\section{Crossing Relations of Types, I, II, III}
\setcounter{equation}{0}
\noindent

The crossing relations used in Sect.5 and partialy in Sect.4 are derived here. 
We use two product expansions:
\bba
    && Y^*_{\half M} Y^{ij}_{J_1(M_1x_1)}
             \nonumber  \\
    && = \fr{1}{\sq{{\rm Vol}(S^3)}} \sum_{I=J_1 \pm \half} \sum_{T,x}
       \eps_T \E^{\half M}_{J_1(M_1x_1), I(-Tx)} Y^{ij}_{I(Tx)}
             \nonumber \\
    && \quad
          - \fr{1}{\sq{{\rm Vol}(S^3)}} \sum_{I=J_1} \sum_{V,y}
          \fr{2}{(2I-1)(2I+3)} \eps_V \H^{\half M}_{J_1(M_1x_1); I(-Vy)}
          \hnabla^{(i} Y^{j)}_{I(Vy)}
             \nonumber \\
    &&  \label{product-ST}
\eea
and 
\bba
    && Y^*_{\half M} \hnabla^{(i}Y^{j)}_{J_1(M_1y_1)}
              \nonumber \\
    && = \fr{1}{\sq{{\rm Vol}(S^3)}} \sum_{I=J_1} \sum_{T,x}
       \eps_T \H^{\half M}_{I(-Tx); J_1(M_1y_1)} Y^{ij}_{I(Tx)}
               \nonumber  \\
    && \quad
          - \fr{1}{\sq{{\rm Vol}(S^3)}} \sum_{I=J_1\pm\half} \sum_{V,y}
          \fr{2}{(2I-1)(2I+3)} \eps_V {}^0\oD^{\half M}_{J_1(M_1y_1), I(-Vy)}
          \hnabla^{(i} Y^{j)}_{I(Vy)}
                \nonumber \\
    && \quad
         +\fr{1}{\sq{{\rm Vol}(S^3)}} \sum_{I=J_1} \sum_S
          \fr{3}{2} \fr{1}{(2I-1)2I(2I+2)(2I+3)}
               \nonumber \\
    && \qquad\qquad\qquad\qquad \times
          \eps_S {}^0\oG^{\half M}_{J_1(M_1y_1); I-S}
          \left( \hnabla^{(i} \hnabla^{j)}
                  -\fr{1}{3} \hgm^{ij} \hnabla^2 \right) Y_{IS}.
                 \label{product-SV}
\eea
Here, note that ${}^0\oD^{\half}_{J_1,J_2}$ and ${}^0\oG^{\half}_{J_1,J_2}$ can be
expressed by $\D^{\half}_{J_1,J_2}$ and $\G^{\half}_{J_1,J_2}$
as (\ref{0D1/2}) and (\ref{0G}), respectively.
The sum of $I$ in each line of r.h.s. is fixed by the triangular conditions
of the Clebsch-Gordan coefficients.
Especially, $\H^\half_{J_1; J_2} \propto \dl_{J_1J_2}$
and $\G^\half_{J_1; J_2} \propto \dl_{J_1J_2}$ are taken into account.

\paragraph{Crossing relations of type I}

We first consider the following series of integrals:
\bba
   n=0 && \quad  \int_{S^3} d\Om_3 Y^*_{\half M} Y^{ij}_{J_1 (M_1x_1)}
             Y_{ij J_2 (M_2x_2)} Y^*_{JM},
                   \nonumber     \\
   n=1 && \quad  \int_{S^3} d\Om_3 Y^*_{\half M} Y^{ij}_{J_1 (M_1x_1)}
             Y^k_{j J_2 (M_2x_2)} \hnabla_{(i}Y^*_{k) J(My)}
                    \nonumber    \\
   n=2 && \quad  \int_{S^3} d\Om_3 Y^*_{\half M} Y^{ij}_{J_1 (M_1x_1)}
             Y^k_{j J_2 (M_2x_2)} Y^*_{ik J(Mx)}
                    \nonumber    \\
   n=3 && \quad  \int_{S^3} d\Om_3 Y^*_{\half M} Y^{ij}_{J_1 (M_1x_1)}
             Y^{kl}_{J_2 (M_2x_2)} \hnabla_{(i} Y^*_{jkl) J(Mz)},
                    \nonumber    \\
   n=4 && \quad  \int_{S^3} d\Om_3 Y^*_{\half M} Y^{ij}_{J_1 (M_1x_1)}
             Y^{kl}_{J_2 (M_2x_2)} Y^*_{ijkl J(Mw)},
\eea
where $n$ denotes the rank of the last harmonics in each integrand. 
Using the product expansion (\ref{product-ST}), we obtain the crossing relation
\bba
  && \sum_{I=J_1 \pm\half} \sum_{T,x} \eps_T \E^{\half M}_{J_1(M_1x_1), I(-Tx)}
             {}^n\oE^{J(M\eps_n)}_{I(Tx), J_2(M_2x_2)}
             \nonumber \\
  && -\sum_{I=J_1} \sum_{V,y} \fr{2}{(2I-1)(2I+3)} \eps_V
          \H^{\half M}_{J_1(M_1x_1); I(-Vy)} 
          {}^n\oH^{J(M\eps_n)}_{J_2(M_2x_2);I(Vy)}
             \nonumber  \\
  && = [ (J_1, M_1, x_1) \leftrightarrow (J_2, M_2, x_2)],
            \label{cross-I}
\eea
for each $n$. These equations refer to the crossing relations of type I.

\paragraph{Crossing relations of type II}

Next, we consider the following series of integrals:
\bba
   n=0 && \quad  \int_{S^3} d\Om_3 Y^*_{\half M} Y^{ij}_{J_1 (M_1x_1)}
             \hnabla_{(i}Y_{j) J_2 (M_2y_2)} Y^*_{JM},
                   \nonumber     \\
   n=1 && \quad  \int_{S^3} d\Om_3 Y^*_{\half M} Y^i_{j J_1 (M_1x_1)}
             \hnabla^{(k} Y^{j)}_{J_2 (M_2y_2)} \hnabla_{(i}Y^*_{k) J(My)}
                    \nonumber    \\
   n=2 && \quad  \int_{S^3} d\Om_3 Y^*_{\half M} Y^i_{j J_1 (M_1x_1)}
             \hnabla^{(k} Y^{j)}_{J_2 (M_2y_2)} Y^*_{ik J(Mx)}
                    \nonumber    \\
   n=3 && \quad  \int_{S^3} d\Om_3 Y^*_{\half M} Y^{ij}_{J_1 (M_1x_1)}
            \hnabla^{(k} Y^{l)}_{J_2 (M_2y_2)} \hnabla_{(i} Y^*_{jkl) J(Mz)},
                    \nonumber    \\
   n=4 && \quad  \int_{S^3} d\Om_3 Y^*_{\half M} Y^{ij}_{J_1 (M_1x_1)}
             \hnabla^{(k} Y^{l)}_{J_2 (M_2y_2)} Y^*_{ijkl J(Mw)}.
\eea
The only difference from the case of type I is that 
$Y^{ij}_{J_2(M_2x_2)}$ is replaced by $\hnabla^{(i} Y^{j)}_{J_2(M_2y_2)}$.
Using the product expansions (\ref{product-ST}) and (\ref{product-SV}),
we obtain the crossing relation
\bba
  && \sum_{I=J_1 \pm\half} \sum_{T,x} \eps_T \E^{\half M}_{J_1(M_1x_1), I(-Tx)}
             {}^n\oH^{J(M\eps_n)}_{I(Tx); J_2(M_2y_2)}
             \nonumber \\
  && -\sum_{I=J_1} \sum_{V,y} \fr{2}{(2I-1)(2I+3)} \eps_V
          \H^{\half M}_{J_1(M_1x_1); I(-Vy)} 
          {}^n\oD^{J(M\eps_n)}_{I(Vy), J_2(M_2y_2)}
             \nonumber  \\
  && = \sum_{I=J_1} \sum_{T,x} \eps_T \H^{\half M}_{I(-Tx); J_2(M_2y_2)}
             {}^n\oE^{J(M\eps_n)}_{I(Tx), J_1(M_1x_1)}
             \nonumber \\
  && -\sum_{I=J_2 \pm\half} \sum_{V,y} \fr{2}{(2I-1)(2I+3)}
            \left\{ I(I+1)+J_2(J_2+1)-\fr{11}{4} \right\}
              \nonumber \\
  && \qquad\qquad\qquad\qquad \times
        \eps_V \D^{\half M}_{J_2(M_2y_2), I(-Vy)} 
        {}^n\oH^{J(M\eps_n)}_{J_1(M_1x_1);I(Vy)}
              \nonumber \\
  && + \sum_{I=J_2} \sum_S  \fr{3}{(2I-1)2I(2I+2)(2I+3)} 
           \left\{ I(I+1)+J_2(J_2+1)-\fr{3}{2} \right\} 
             \nonumber \\
  && \qquad\qquad\qquad\qquad \times 
     \eps_S \G^{\half M}_{J_2(M_2y_2);I -S} 
         {}^n\oI^{J(M\eps_n)}_{J_1(M_1x_1); I S},
            \label{cross-II}
\eea
for each $n$. These equation refer to the crossing relations of type II.

\paragraph{Crossing relations of type III}

Finally, we consider the following series of integrals:
\bba
   n=0 && \quad  \int_{S^3} d\Om_3 Y^*_{\half M} \hnabla^{(i} Y^{j)}_{J_1 (M_1y_1)}
             \hnabla_{(i}Y_{j) J_2 (M_2y_2)} Y^*_{JM},
                   \nonumber     \\
   n=1 && \quad  \int_{S^3} d\Om_3 Y^*_{\half M} \hnabla^{(i} Y^{k)}_{J_1 (M_1y_1)}
             \hnabla^{(j} Y_{k) J_2 (M_2y_2)} \hnabla_{(i}Y^*_{j) J(My)}
                    \nonumber    \\
   n=2 && \quad  \int_{S^3} d\Om_3 Y^*_{\half M} \hnabla^{(i} Y^{k)}_{J_1 (M_1y_1)}
             \hnabla^{(j} Y_{k) J_2 (M_2y_2)} Y^*_{ij J(Mx)}
                    \nonumber    \\
   n=3 && \quad  \int_{S^3} d\Om_3 Y^*_{\half M} \hnabla^{(i} Y^{j)}_{J_1 (M_1y_1)}
            \hnabla^{(k} Y^{l)}_{J_2 (M_2y_2)} \hnabla_{(i} Y^*_{jkl) J(Mz)},
                    \nonumber    \\
   n=4 && \quad  \int_{S^3} d\Om_3 Y^*_{\half M} \hnabla^{(i} Y^{j)}_{J_1 (M_1y_1)}
             \hnabla^{(k} Y^{l)}_{J_2 (M_2y_2)} Y^*_{ijkl J(Mw)}.
\eea
The difference from the cases of type I is that $Y^{ij}_{J_1(M_1x_1)}$ 
and $Y^{ij}_{J_2(M_2x_2)}$ are replaced by $\hnabla^{(i} Y^{j)}_{J_1(M_1y_1)}$ 
and $\hnabla^{(i} Y^{j)}_{J_2(M_2y_2)}$, respectively. 
Using the product expansion (\ref{product-SV}),
we obtain the crossing relation
\bba
  && \sum_{I=J_1} \sum_{T,x} \eps_T \H^{\half M}_{I(-Tx);J_1(M_1y_1)}
             {}^n\oH^{J(M\eps_n)}_{I(Tx); J_2(M_2y_2)}
             \nonumber \\
  && -\sum_{I=J_1 \pm\half} \sum_{V,y} \fr{2}{(2I-1)(2I+3)}
         \left\{ I(I+1)+J_1(J_1+1)-\fr{11}{4} \right\}
             \nonumber \\
  && \qquad\qquad\qquad\qquad \times
        \eps_V \D^{\half M}_{J_1(M_1y_1); I(-Vy)}
          {}^n\oD^{J(M\eps_n)}_{I(Vy), J_2(M_2y_2)}
             \nonumber  \\
  && + \sum_{I=J_1} \sum_S  \fr{3}{(2I-1)2I(2I+2)(2I+3)} 
           \left\{ I(I+1)+J_1(J_1+1)-\fr{3}{2} \right\}
             \nonumber \\
  && \qquad\qquad\qquad\qquad \times 
        \eps_S \G^{\half M}_{J_1(M_1y_1);I -S} 
          {}^n\oG^{J(M\eps_n)}_{J_2(M_2y_2);I S}
             \nonumber \\
  && = [(J_1, M_1, y_1) \leftrightarrow (J_2, M_2, y_2) ],
              \label{cross-III}
\eea
for each $n$. These equations refer to the crossing relations of type III.

\section{Conformal Algebra in Ghost Sector}
\setcounter{equation}{0}
\noindent

   In this section we reinvestigate conformal algebra in the ghost sector 
discussed in Ref.\cite{amm97}  
As discussed in the previous paper~\cite{hh}, in the radiation$^+$ gauge, 
there is the residual gauge symmetry generated by the conformal Killing vectors. 
The 15 ghosts to fix it are described as $\bfc$, $\bfc_M$, $\bfc_M^\dag$ and $\bfc_{MN}$, 
where the indices, $M$ and $N$, indicate the $({\bf 2},{\bf 2})$ representation 
of $SU(2)\times SU(2)$, and $\bfc_{MN}$ satisfies the
equations $\bfc^\dag_{MN}=\bfc_{NM}$ and $\bfc_{MN}=-\eps_M\eps_N\bfc_{-N-M}$. 
We also intorduce associate anti-ghost fields, denoted 
by $\bfb$, $\bfb_M$, $\bfb_M^\dag$ and $\bfb_{MN}$, 
where $\bfb^\dag_{MN}=\bfb_{NM}$ and $\bfb_{MN}=-\eps_M\eps_N\bfb_{-N-M}$. 
The commutators of these fields are defined as
\bba
   \{ \bfb,  \bfc \} &=& 1, 
         \nonumber \\
   \left\{ \bfb^\dag_M, \bfc_N \right\} &=& 
     \left\{ \bfb_M, \bfc^\dag_N \right\} = \dl_{MN},
         \nonumber \\
   \left\{ \bfb_{M_1N_1}, \bfc_{M_2N_2} \right\} &=& 
     \dl_{M_1M_2}\dl_{N_1N_2}-\eps_{M_1}\eps_{N_1}\dl_{-M_1N_2}\dl_{-N_1M_2}.  
\eea
The ghosts related to the time translation, $\bfc$ and $\bfb$, 
and the rotations on $S^3$, $\bfc_{MN}$ and $\bfb_{MN}$, have the level $0$. 
The ghosts related  to the special conformal transformations, $\bfc_M$ and $\bfb_M$, 
have the level $-1$, and their conjugates, $\bfc^\dag_M$ and $\bfb^\dag_M$, have the level $1$.  
Thus, the hamiltonian has the form 
\bb
     H = \sum_R \left( \bfb^\dag_R \bfc_R +\bfc^\dag_R \bfb_R \right) 
         + \hbox{constant}.
\ee

    Since the conformal charge $Q_M$ has the level $-1$ and belongs to 
the $({\bf 2},{\bf 2})$ representation of $SU(2)\times SU(2)$, the general form of this charge 
is given by
\bb
   Q_M = \lam_1 \bfb \bfc_M +\lam_2 \bfb_M \bfc 
         +\sum_R \left( \kappa_1 \bfb_R \bfc_{RM} 
                        + \kappa_2 \bfb_{MR}\bfc_R \right).  
\ee  
We here require that these charges form the conformal algebra (\ref{algebra}). 
This requirement is satisfied if $\lam_1 \lam_2 =2$ and $\kappa_1 \kappa_2=2$,    
and thus we obtain the following $15$ conformal charges: 
\bba
      Q_M &=& \lam \bfb \bfc_M +\fr{2}{\lam} \bfb_M \bfc 
         +\sum_R \left( \kappa \bfb_R \bfc_{RM} 
                        + \fr{2}{\kappa} \bfb_{MR}\bfc_R \right),
           \nonumber \\
      H &=& \sum_R \left( \bfb^\dag_R \bfc_R +\bfc^\dag_R \bfb_R \right) -4, 
           \nonumber \\ 
      R_{MN} &=& \bfb^\dag_N \bfc_M +\bfc^\dag_N \bfb_M 
                  -\eps_M\eps_N \left( \bfb^\dag_{-M} \bfc_{-N} 
                       + \bfc^\dag_{-M} \bfb_{-N} \right) 
                \nonumber  \\
             && -\half\sum_R \left( \bfb_{MR}\bfc_{NR} +\bfc_{RM}\bfb_{RN} 
                     -\bfb_{RN}\bfc_{RM} -\bfc_{NR}\bfb_{MR} \right),
\eea
where $\lam$ and $\kappa$ are arbitrary constants. Note that the constant term 
in the Hamiltonian is fixed to be $-4$. This constant has a relationship to the world-volume 
dimensions.

\end{document}